\documentclass[12pt]{article}
\usepackage[english]{babel}
\usepackage{amsthm,amsmath,amssymb, amsbsy,amsfonts,latexsym,euscript}

     \title{Self-adjoint elliptic problems in domains with cylindrical ends under weak assumptions on the stabilization of coefficients}
     \author{Kalvine V.\,O.}

   \date{}
  \newtheorem{thm}{Theorem}[section]
  \newtheorem{prop}[thm]{Proposition}
  \newtheorem{lm}[thm]{Lemma}
  
  \newtheorem{nb}[thm]{Remark}
  \newtheorem{defn}[thm]{Definition}

\makeatletter \@addtoreset{equation}{section}

\begin{document}\maketitle
\begin{abstract}
The general self-adjoint elliptic boundary value problems are
considered in a domain $G\subset \Bbb R^{n+1}$ with finitely many
cylindrical ends. The coefficients are stabilizing (as
$x\to\infty$, $x\in G$) so slowly that we can only describe some
``structure'' of solutions far from the origin. This problem may
be  understood as a model of ``generalized branching waveguide.''
We introduce a notion of the energy flow through the
cross-sections of the cylindrical ends and define outgoing and
incoming ``waves.'' An augmented scattering matrix is introduced.
Analyzing the spectrum of this matrix one can find the number of
linearly independent solutions to the homogeneous problem
decreasing at infinity with a given rate. We discuss the statement
of problem with so-called radiation conditions and
 enumerate self-adjoint extensions of the operator of the problem.
\end{abstract}

\section{Introduction}
In domain $G\subset \Bbb R^{n+1}$ with finitely many cylindrical
ends we consider the general formally self-adjoint boundary value
problem. The coefficients tend to limits (as $x\to\infty$, $x\in
G$) too slow to allow obtaining an asymptotic of solution at
infinity. Using the results of the paper \cite{1} (see also
\cite[Section 8.5]{2}), one can get some ``structure'' of solution
to the problem: far from the origin a solution is represented as a
linear combination of some functional series plus a remainder. The
coefficients in the linear combination remain unknown. In this
paper we develop an approach, which, in particular,  allows  to
derive expressions for the coefficients in the structure of
solution to the problem under consideration.

  Let $\Pi^r_+=\{(y,t): y\in \Omega^r,
t>0\}$, $r=1,\dots, N$, stand for the cylindrical ends, where
$\Omega^r$ is the cross-section. (The domain $G$ coincides with
the union $\Pi^1_+\cup \dots \cup \Pi^N_+$ outside a large ball.)
With every cylindrical end $\Pi^r_+$ we associate limit and model
problems in the cylinder $\Pi^r=\{(y,t): y\in \Omega^r, t\in \Bbb
R\}$.

 As the coefficients of limit problem we take the limits of
coefficients of the original problem as $t\to +\infty$, $(y,t)\in
\Pi^r_+$. It is assumed that the limit problems are elliptic.
Since the operator of the original problem is formally
self-adjoint, the operators of the limit problems are formally
self-adjoint as well. As is known (see e.g. \cite[Chapter 5]{3}),
one can consider every limit problem  as a model of ``generalized
waveguide.'' This means that a generalized notion of the energy
flow through the cross-section of the cylinder is introduced, the
solution to the homogeneous problem is called incoming (outgoing)
wave if the energy flow associated with the solution is positive
(negative). The amplitudes of such waves may grow with power or
even with exponential rate at infinity.

The operator of the model problem is formally self-adjoint and
depends on the parameter $T\in \Bbb R$. The coefficients of the
model problem coincide with the coefficients of the original
problem on the set $\{(y,t)\in  P^r_+,t>T+3\}$ and with their
limits (as $t\to +\infty$, $(y,t)\in \Pi^r_+$) on the set
$\{(y,t)\in \Pi^r,t<T\}$. The coefficients of the model problem
tend to the coefficients of the limit one as $T\to+\infty$. Thus a
solution to the homogeneous model problem can be obtained in the
form of functional series by the method of successive
approximations, as the first approximation it is natural to take a
wave of the limit problem. On the analogy of the limit problem,
for the model problem we introduce a notion of the energy flow
through the cross-section of the cylinder. The formula for the
energy flow through the cross-section $\{(y,t)\in \Pi^r, t=R\}$,
$R<T$, is the same for both (limit and model) problems because the
coefficients of the problems coincide on the set $\{(y,t)\in
\Pi^r, t<T\}$. Moreover, it turns out that a wave and the
correspondent solution to the homogeneous model problem have equal
energy flows through the left infinitely distant cross-section of
$\Pi^r$. This allows to calculate the energy flows of obtained
solutions to the homogeneous model problem and also allows to
separate these solutions into incoming and outgoing waves (of the
model problem). Due to the formally self-adjointness of the model
problem operator, the energy flows of such waves remain constant
along the cylinder. Recall that the coefficients of the model
problem coincide with the coefficients of original problem on the
set $\{(y,t)\in \Pi^r_+, t>T+3\}$. Owing to this fact, one can
consider the domain $G$ as a branching waveguide, where the waves
obtained for the model problem in $\Pi^r$ propagate along the
cylindrical end $\Pi^r_+$ of $G$, $r=1,\dots,N$. Using a
modification of the scheme suggested in \cite[Theorem 6.2]{1}, we
get the structure of solutions to the problem in $G$:  far from
the origin a solution is represented as a linear combination of
the waves plus a remainder. Some waves properties obtained on the
previous step allow us to derive the formulas for the coefficients
in the structure of solution. The results are represented in
Theorem~\ref{pasv} and Theorem~\ref{taswG}.

The remaining part of the paper basically contains corollaries of
the theorems~\ref{pasv} and~\ref{taswG}.  We omit the proofs
because they almost repeat the proofs of the similar assertions in
\cite[Chapther 5]{3} or in \cite{11}, where it is assumed that the
coefficients are stabilizing with exponential rate. The changes in
the proofs mainly consist in usage of Theorem~\ref{pasv} or
Theorem~\ref{taswG} instead of asymptotic representations. In the
main text we insert the exact references to the needed proofs.

 The operator of the
problem acts in weighted spaces. We obtain some information about
the kernel of the problem (Propositions \ref{cl2} and \ref{cl5})
and introduce ``scattering matrices.'' These unitary matrices take
into account waves growing at infinity. Analyzing the spectrum of
this matrix one can find the number of linearly independent
solutions to the homogeneous problem decreasing at infinity with a
given rate (cf. Proposition~\ref{crit}). We discuss the statement
of problem with ``radiation conditions:'' the domain of operator
contains only functions with prescribed structure at infinity.
This is a way to choose a solution (with a certain arbitrariness)
(cf. Propositions \ref{cl3} and \ref{cl4}). The intrinsic
radiation conditions (the solution mainly consists of outgoing
waves) can be utilized in every case. To verify whether given
radiation conditions can be used, it is required to know the
scattering matrix (cf. Proposition \ref{cl6}).
 In Section~\ref{sss} the self-adjoint extensions of operator of the problem are found.

Some of the results proved in our paper were announced earlier in
the work \cite{4}.

\section{Statement of the problem and preliminaries}
\subsection{Domain and self-adjoint boundary value problem}
Let $G\subset \Bbb R^{n+1}$ be a domain with smooth boundary
$\partial G$ coinciding, outside a large ball, with the union
$\Pi^1_+\cup \dots \cup \Pi^N_+$ of non-overlapping semicylinders;
here $\Pi^r=\{(y^r,t^r): y^r\in\Omega^r, t^r>0 \}$, $(y^r,t^r)$
 are local coordinates, and the cross-section $\Omega^r$ is bounded domain in $\Bbb R^n$.
In the domain $G$ we introduce a formally self-adjoint $k\times
k$-matrix $\mathcal
 L$ of differential operators $\mathcal L_{ij}(x,D_x)$  with smooth coefficients,
 where ${\rm ord \,} \mathcal L_{ij}=\tau_i+\tau_j$, the numbers $\tau_1,\dots,\tau_k$
 are non-negative integers, and $\tau_1+\cdots+\tau_k=m$. Consider the boundary value problem
\begin{equation}\label{1}
\begin{split}
{\mathcal L} (x,D_x)u(x)&=f(x),\qquad x\in G, \\
{\mathcal B}  (x,D_x)u(x)&=g(x),\qquad x\in\partial G,
\end{split}
\end{equation}
where $\mathcal B$ is an $m\times k$-matrix  of differential
operators. For a given $\mathcal L$ we find a class of boundary
conditions such that for an element $\mathcal B$ of the class the
self-adjoint Green formula
\begin{equation}\label{1a}
(\mathcal L u, v)_G+(\mathcal B u,\mathcal Q v)_{\partial
G}=(u,\mathcal L v)_G+(\mathcal Q u,\mathcal B v)_{\partial G}
\end{equation}
holds with some $m\times k$-matrix $\mathcal Q$ of differential
operators for all $u,v\in C_c^\infty(\overline G)$. It is supposed
 that $\mathcal B$ in (\ref{1}) is from the mentioned class and the
 problem is elliptic.

\subsection{Boundary conditions and self-adjoint Green
formula}\label{s2}
 If necessary changing the enumeration of the rows and columns in $\mathcal
 L_{ij}(x,D_x)$, we may always arrange that
 $\tau=\tau_1\geq\tau_2\geq\dots\geq\tau_k$. In what follows we
 suppose that this has been done. Denote by $K_s$,
 $s=1,\dots,\tau$, the number of values $j$ such that
 $\tau_j\geq\tau-s+1$. Then $K_1+\dots +K_\tau=\tau_1+\dots+\tau_k=m$ and
  $ K_1\leq\dots\leq K_\tau \leq\ k$.  On
 the boundary $\partial G$ we introduce the $m\times k$-matrix
 \begin{equation}\label{D}
 \mathcal D=\left(%
\begin{array}{c}
{\bf D}^1 \\
  \vdots \\
  {\bf D}^\tau \\
\end{array}%
\right),
 \end{equation}
 where the block ${\bf D}^s$ consists of the rows
 $$
 (\delta_{1,h},\dots,\delta_{k,h})D_\nu^{\tau_h-\tau+s-1},\qquad
 h=1,\dots,K_s;
 $$
 here $\nu$ is the unit outward normal to $\partial G$ and
 $D_\nu=-i\partial/\partial \nu$. With $\mathcal L(x,D_x)$
 we associate the sesquilinear form
\begin{equation}\label{2}
a(u,v)=\sum_{i,j=1}^k \sum_{|\eta|\leq\tau_h}\sum_{|\mu|\leq
\tau_j}\int_G a_{ij}^{\eta\mu}(x) D^\mu_x u_j(x)\overline{D^\eta_x
v_i(x)}\,dx.
\end{equation}
The Green formula
\begin{equation}\label{GF}
a(u,v)=(\mathcal L u,v)_G+(\mathcal N u,\mathcal D v)_{\partial G}
\end{equation}
holds for $u,v\in C_c^\infty (\overline{ G})$ with some $m\times
k$-matrix $\mathcal N (x,D_x)=\|\mathcal N_{qj} (x,D_x)\|$ of
differential operators, $\operatorname{ord} \mathcal
N_{qj}+\operatorname{ord}\mathcal D_{qh}\leq\tau_h+\tau_j-1$;  by
$(\cdot,\cdot)_G$ and $(\cdot,\cdot)_{\partial G}$ we denote the
inner products on $L_2(G)$ and $L_2(\partial G)$.
\begin{nb}\label{ip}{\rm
The proof of the Green formula (\ref{GF}) is standard. Using
``local maps,''  we can consider $G$ as the half-space $\Bbb
R^n_-=\{x\in\Bbb R^n, x_n<0 \}$ (cf. \cite{5}, \cite{6}). Then
$D_\nu=D_{x_n}$ and $\mathcal
L_{ij}(x,D_x)=\sum_{|\alpha|+\beta\leq\tau_i}\sum_{|\mu|\leq\tau_j}D_\tau^\alpha
D_\nu^\beta a_{ij}^{(\alpha\beta)\mu}(x)D_x^\mu$, where
$D_\tau=D_{x_1}D_{x_2}\dots D_{x_{n-1}}$. Integrating by parts and
changing the order of summation, we obtain
\begin{equation*}
\begin{split}
&(\mathcal L u,
v)_G=\sum_{i,j=1}^k\sum_{|\alpha|+\beta\leq\tau_i}\sum_{|\mu|\leq
\tau_j}\bigl(D_\tau^\alpha D_\nu^\beta
a_{ij}^{(\alpha\beta)\mu}(x)D_x^\mu u_j,
v_i\bigr)_G=a(u,v)\\&+\sum_{s=1}^\tau\sum_{i=1}^{K_s}\sum_{j=1}^k\sum_{|\mu|\leq\tau_j}\sum_{|\alpha|+\beta\leq\tau_j,\beta\geq
s}\bigl(D_\tau^\alpha D_\nu^{\beta-s+\tau-\tau_i}
a_{ij}^{(\alpha\beta)\mu}(x)D_x^\mu u_j, D_\nu^{\tau_i -\tau +s-1}
v_i \bigr)_{\partial G},
\end{split}
\end{equation*}
where $u,v\in C_c^\infty(\overline G)$. This implies (\ref{GF}).}
\end{nb}

\begin{defn}\label{R}
A matrix $\mathcal P=\mathcal P(x,D_x)$ is called a Dirichlet
system on the boundary $\partial G$ if there exists an $m\times
m$-matrix $\mathcal R=\mathcal R(x,D_x)$ satisfying the following
conditions.

(i) $\mathcal P(x,D_x)=\mathcal R(x,D_x)\mathcal D(x,D_x)$, where
$\mathcal D$ is given in (\ref{D}).

(ii) The matrix $\mathcal R$ consists of $K_p\times K_s$-blocks
${\mathcal  R}_{[p,s]}$ ($p,s=1,\dots,\tau$). The elements of
${\mathcal R}_{[p,s]}$, $s\leq p$, are tangential differential
operators with smooth coefficients on $\partial G$ of order not
higher than $p-s$, while the elements of ${\mathcal  R}_{[p,s]}$,
$s>p$, are zeros. The ${\mathcal  R}_{[p,p]}(x)$ are nondegenerate
matrices, $|\det{\mathcal  R}_{[p,p]}(x)|>\varepsilon>0$ for
$x\in\partial G$.

\end{defn}

If in particular $k=1$, then $\mathcal P(x,D_x)$ is the usual
Dirichlet system of order $\tau$ (see \cite{7},\cite{8},\cite{9}).

\begin{nb}\label{Rr} The operator $\mathcal R$ from Definition~\ref{R}
has an inverse $\mathcal R^{-1}$, which is a matrix of
differential operators of the same structure as $\mathcal R$. The
block ${\mathcal  R}^{-1}_{[p,p]}(x)$ of $\mathcal R^{-1}(x,D_x)$
is the inverse matrix $({\mathcal R}_{[p,p]}(x))^{-1}$,
$p=1,\dots,\tau$. The blocks ${\mathcal R}^{-1}_{[p,s]}$, $s<p$,
of sizes $K_p\times K_s$ can be successively found from the
relations
$$
{\mathcal  R}^{-1}_{[p,s]}(x,D_x)={\mathcal
R}_{[p,p]}^{-1}(x)\sum_{\ell=s}^{p-1}(-{\mathcal
R}_{[p,\ell]}(x,D_x)){\mathcal R}_{[\ell, s]}^{-1}(x,D_x).
$$
The blocks ${\mathcal R}^{-1}_{[p,s]}$, $s>p$, consist of zeros.
\end{nb}
Let
\begin{equation}\label{p}
\mathcal P(x,D_x)=\mathcal R(x,D_x)\mathcal D(x,D_x)
\end{equation}
 be a Dirichlet system on $\partial G$. We set
\begin{equation}\label{t}
 \mathcal T(x,D_x)={\mathcal R}^{-1}_*(x,D_x)\mathcal
N(x,D_x),
\end{equation}
 where ${\mathcal R}^{-1}_*(x,D_x)$ is the formally
adjoint differential operator to ${\mathcal R}^{-1}(x,D_x)$ and
$\mathcal N(x,D_x)$ is from the Green formula (\ref{GF}).
Introduce $m\times k$-matrices $\mathcal B$ and $\mathcal Q$ such
that
\begin{equation}\label{3}
(\mathcal B_{q1},\dots,\mathcal B_{qk})=(\mathcal
T_{q1},\dots,\mathcal T_{qk}), \quad (\mathcal
Q_{q1},\dots,\mathcal Q_{qk})=(\mathcal P_{q1},\dots,\mathcal
P_{qk})
\end{equation}
 for some numbers $q$, $1\leq q\leq m$, while
\begin{equation}\label{4}
(\mathcal B_{q1},\dots,\mathcal B_{qk})=(\mathcal
P_{q1},\dots,\mathcal P_{qk}), \quad (\mathcal
Q_{q1},\dots,\mathcal Q_{qk})=-(\mathcal T_{q1},\dots,\mathcal
T_{qk})
\end{equation}
for the remaining rows of $\mathcal B$ and $\mathcal Q$.
Therefore,
\begin{equation}\label{i.1}
\begin{split}
(\mathcal N u,\mathcal D v)_{\partial G}-(\mathcal D u,\mathcal N
v)_{\partial G}=(\mathcal N u,\mathcal R^{-1}\mathcal P
v)_{\partial G}-(\mathcal R^{-1}\mathcal P u,\mathcal N
v)_{\partial G}\\=(\mathcal Tu,\mathcal P v)_{\partial
G}-(\mathcal P u,\mathcal T v)_{\mathcal G}=(\mathcal B u,\mathcal
Q v)_{\partial G}-(\mathcal Q u,\mathcal B v)_{\partial G}.
\end{split}
\end{equation}
Since $\mathcal L$ is formally self-adjoint, the form (\ref{2}) is
symmetric (i.e. $a(u,v)=\overline {a(v,u)}$).  From (\ref{GF})  we
obtain
\begin{equation}\label{i.2}
(\mathcal L u,v)_G+(\mathcal N u, \mathcal D v)_{\partial
G}=(u,\mathcal L v)_G+(\mathcal D u,\mathcal N v)_{\partial G}.
\end{equation}
Together with (\ref{i.1}) this leads to the Green formula
(\ref{1a}).
\subsection{Limit operators}\label{slo}
Let $r$, $r=1,\dots,N$, be a fixed number. We write the
superscript $r$ at $\mathcal L$, $\mathcal R$, and other operators
if they are written in the coordinates $(y,t)$ inside the
semicylinder $\overline\Pi^r_+$. Let $\mathcal L^r=\|\mathcal
L^r_{ij}\|$ and let
\begin{equation}\label{i.3a}
\mathcal
L^r_{ij}(y,t,D_y,D_t)=\sum_{|\eta|+\mu\leq\tau_j+\tau_h}\ell_{ij}^{\eta\mu}(y,t)D^\eta_y
D^\mu_t.
\end{equation}
We set $\psi_T(y,t)\equiv\psi(t-T)$ for
$(y,t)\in\overline\Pi^r_+$, where $\psi\in C^\infty(\Bbb R)$ is a
cutoff function such that $\psi(t)=0$ for $t<1$ and $\psi(t)=1$
for $t>2$.
\begin{defn}
We say that $\mathcal L$ is stabilizing in the semicylinder
$\Pi^r_+$ if there exist functions ${\bf l}_{hj}^{\eta\mu}$ of
$y\in\overline \Omega^r$ ({\it limit coefficients}) such that
\begin{equation}\label{i.4}
\lim_{T\to +\infty}\|\psi_T({\ell}_{ij}^{\eta\mu}-{\bf
l}_{ij}^{\eta\mu});C^\infty(\overline\Pi^r_+)\|=0,\quad
|\eta|+\mu\leq\tau_i+\tau_j,\quad i,j=1,\dots,k.
\end{equation}
\end{defn}

\begin{defn}
Let $\mathcal F(x,D_x)=\|\mathcal F_{qj}(x,D_x)\|$ be an operator
given on the boundary $\partial G$. Write down $\mathcal F_{qh}$
in the local coordinates (y,t):
\begin{equation*}
\mathcal F^r_{q
j}(y,t,D_y,D_t)=\sum_{|\alpha|+\beta\leq\operatorname{ord}\mathcal
F_{qj}} f_{qj}^{\alpha \beta}(y,t)D_y^\alpha D^\beta_t.
\end{equation*}
We say that $\mathcal F$ is stabilizing in $\Pi^r_+$ if there
exist functions ${\bf f}_{qj}^{\alpha \beta}$ of $y\in\partial
\Omega^r$ such that
\begin{equation*}
\lim_{T\to+\infty}\|\psi_T(f_{qj}^{\alpha\beta}-{\bf
f}_{qj}^{\alpha\beta});C^\infty(\partial\Omega^r\times \Bbb
R_+)\|=0
\end{equation*}
for $|\alpha|+\beta\leq\operatorname{ord}\mathcal F_{qj}$ and for
all values of $q$ and $j$; here $\Bbb R_+=\{t\in\Bbb R: t>0\}$.
\end{defn}
Since the coefficients of $\mathcal D_{qj}^r$ do not depend on
$t$, the operator $\mathcal D(x,D_x)$ from (\ref{D}) is
stabilizing in $\Pi^1_+,\dots,\Pi^N_+$. Assume that $\mathcal L$
is stabilizing in $\Pi^r_+$. The stabilization of $\mathcal
N(x,D_x)$ in $\Pi^r_+$ is guaranteed by (\ref{i.4}) (the
coefficients of $\mathcal N^r(y,t,D_y,D_t)$ are expressed in terms
of
 $\ell_{ij}^{\eta\mu}$; see Remark \ref{ip}).
Therefore an operator $\mathcal B(x,D_x)$ constructed of the rows
of $\mathcal N (x, D_x)$ and $\mathcal D (x,D_x)$ is stabilizing
in $\Pi^r$ as well; this case $\mathcal R(x,D_x)\equiv I$, see
(\ref{p}), (\ref{t}) and (\ref{3}), (\ref{4}). In the general case
we assume the stabilization of $\mathcal R(x,D_x)=\|\mathcal
R_{hq}(x,D_x)\|_{h,q=1}^m$. Then the operator $\mathcal R^{-1}$ is
stabilizing (see Remark \ref{Rr}), we get the stabilization of the
operators $\mathcal P$, $\mathcal T$, $\mathcal B$, and $\mathcal
Q$.

Let the elements ${  L}^r_{ij}(y,D_y,D_t)$ of the limit operator
${  L}^r=\|{  L}^r_{ij}\|$ be given by the right-hand side of
(\ref{i.3a}) with $\ell_{ij}^{\eta\mu}$ replaced by ${\bf
l}_{ij}^{\eta\mu}$. Likewise, changing the coefficients to the
limit ones, we define the limit operators ${  N}^r$, ${  R}^r$, ${
B}^r$ and etc. The relations ${  P}^r={  R}^r{\mathcal D}^r$ and
${ T^r}=({  R}^r)_*^{-1}{  N}^r$ are fulfilled, where
$({R}^r)_*^{-1}$ is formally adjoint to $({  R}^r)^{-1}$. From
(\ref{3}) and (\ref{4}) it follows that the matrix ${  B}^r$ and
${  Q}^r$ consist of the rows of ${  P}^r$ and ${  T}^r$. The
Green formula
\begin{equation}\label{GFl}
({  L}^r u,v)_{\Pi^r}+({  B}^r u,{  Q}^r v)_{\partial \Pi^r}=(u,{
L}^r v)_{\Pi^r}+({  Q}^r u, {  B}^r v)_{\partial \Pi^r}
\end{equation}
is valid in the cylinder $\Pi^r=\Omega^r\times\Bbb R$, where
$u,v\in C_c^\infty(\overline\Pi^r)$. We assume that the limit
problem
\begin{equation}\label{lp}
\begin{split}
{  L}^r (y,D_y, D_t) u(y,t)&=F(y,t),\quad (y,t)\in\Pi^r,\\
{  B}^r (y,D_y, D_t) u(y,t)&=G(y,t),\quad (y,t)\in\partial\Pi^r,
\end{split}
\end{equation}
is elliptic.

Denote by $W^l_\gamma(\Pi^r)$ the space with norm
$\|e_\gamma\cdot;H^l(\Pi^r)\|$, where $H^l(\Pi^r)$ is the Sobolev
space, $e_\gamma: (y,t)\mapsto \exp(\gamma t)$, and $\gamma\in
\Bbb R$. For $l\geq\tau$ we set
\begin{equation}\label{sp}
\begin{split}
\mathcal D_\gamma^l(\Pi^r) &= \prod_{j=1}^k H_\gamma^{l+\tau_j}(\Pi^r),\\
\mathcal R_\gamma^l(\Pi^r) &=\prod_{i=1}^k
H_\gamma^{l-\tau_i}(\Pi^r)\times \prod_{q=1}^m
H_\gamma^{l-\tau-\sigma_q-1/2}(\partial \Pi^r)
\end{split}
\end{equation}
with $\sigma_q=\operatorname{ord} B^r_{qj}-\tau-\tau_j$,
$\sigma_q<0$.
 The map
 \begin{equation}\label{lo}
 A^r_\gamma=\{L^r,B^r\}:\mathcal D_\gamma^\ell(\Pi^r)\to\mathcal
R_\beta^\ell(\Pi^r)
\end{equation}
is continuous. We introduce the operator pencil
\begin{equation}\label{op}
\Bbb C\ni\lambda\mapsto\mathfrak
A^r(\lambda)=\{L^r(y,D_y,\lambda),B^r(y,D_y,\lambda)\}
\end{equation}
in the domain $\Omega^r$. The spectrum of $\mathfrak A^r$ is
symmetric about the real line and consists of normal eigenvalues.
Any strip $\{\lambda\in\Bbb C:|\operatorname{Im}\lambda|\leq
h<\infty\}$ contains at most finitely many points of the spectrum.
 Denote by
$\lambda_{-\nu^0},\dots,\lambda_{\nu^0}$ with $\nu^0\geq 0$ all
the real eigenvalues of $\mathfrak A^r$ (if the number of real
eigenvalues is even, then $\lambda_0$ is absent). We enumerate the
nonreal eigenvalues so that
$0<\mbox{Im}\,\lambda_{\nu^0+1}\leq\mbox{Im}\,\lambda_{\nu^0+2}\leq\dots$
and $\lambda_\nu=\overline\lambda_{-\nu}$, where
$\nu=\nu^0+1,\nu^0+2,\dots$. Let
$\{\varphi_\nu^{(0,j)},\dots,\varphi_\nu^{(\varkappa_{j\nu}-1,j)};j=1,\dots,J_\nu:=\dim\ker\mathfrak
A^r(\lambda_\nu)\}$ be a canonical system of Jordan chains of the
pencil $\mathfrak A^r$ corresponding to $\lambda_\nu$, i.e.
$\varphi_\nu^{(0,j)}$ is an eigenvector and
$\varphi_\nu^{(1,j)},\dots,\varphi_\nu^{(\varkappa_{j\nu}-1,j)}$
are associated vectors (e.g., see \cite{10}). The functions
\begin{equation}\label{2.11}
     u_\nu^{(\sigma, j)}(y,t)=\exp (i\lambda_\nu t)\sum_{\ell=0}^\sigma
     \frac 1 {\ell !} (it)^\ell \varphi_\nu^{(\sigma-\ell,j)}(y)
\end{equation}
with $\sigma =0,\dots,\varkappa_{j\nu}-1$ satisfy the homogeneous
problem (\ref{lp}). We introduce the form
\begin{equation}\label{2.12}
    q^r(u, v):=({L}^r u,v)_{\Pi^r}+({ B}^r u, { Q}^r v)_{\partial \Pi^r}-
    (u, { L}^r v)_{\Pi^r}-({ Q}^r u, { B}^r v)_{\partial \Pi^r}.
\end{equation}
It is obvious that $q^r(u,v)=-\overline {q^r(v,u)}$ and
$q^r(u,u)\in i \Bbb R$. The Green formula (\ref{GFl}) extends by
continuity to the functions $U\in\mathcal D_\gamma^l(\Pi^r)$ and
$V\in\mathcal D_{-\gamma}^l(\Pi^r)$, therefore $q^r(U,V)=0$.
  \begin{prop}[see \cite{3}]\label{p2.1}
{\rm  (i)} Let $\chi \in C^\infty(\mathbb R)$, $\chi (t)=1$ for
$t\geq2$ and $\chi (t)=0$ for $t\leq 1$. One can choose Jordan
chains $\{\varphi_\nu^{(\sigma,j)}\}$
     to satisfy the following conditions:
\begin{eqnarray}
\label{2.13}q^r(\chi u_\nu^{(\sigma,j)},\chi u_\mu^{(\tau,p)})  =i\delta_{-\nu,\mu}\delta_{j,p}\delta_{\varkappa_{j\nu}-1-\sigma,\tau},\quad |\nu|>\nu_0, |\mu|>\nu_0,\\
\label{2.14} q^r(\chi u_\nu^{(\sigma,j)},\chi
u_\mu^{(\tau,p)})=\pm
i\delta_{\nu,\mu}\delta_{j,p}\delta_{\varkappa_{j\nu}-1-\sigma,\tau},\quad |\nu|\leq \nu_0,|\mu|\leq \nu_0,\\
\label{2.15}q^r(\chi u_\nu^{(\sigma,j)},\chi u_\mu^{(\tau,p)})
=0,\quad |\nu|\leq \nu_0,|\mu|>\nu_0,
\end{eqnarray}
where the functions $u_\nu^{(\sigma,j)}$ are given in
(\ref{2.11}). In (\ref{2.14}) the sign depends on $\nu$ and $j$
(and cannot be taken arbitrarily). The equality (\ref{2.15})
remains true for arbitrary choice of Jordan chains and for any
superscripts. The conditions~(\ref{2.13}) -- (\ref{2.15}) do not
depend on the choice of $\chi$.

\noindent{\rm (ii)} The map (\ref{lo}) is an isomorphism if and
only if the line $\Bbb R+i\gamma=\{\lambda\in \Bbb
C:\operatorname{Im} \lambda=\gamma\}$ is free of the spectrum of
the pencil $\mathfrak A^r$.
\end{prop}

Let $\gamma\geq 0$. Denote by $\mathcal W_\gamma(\Pi^r)$ the
linear span of the functions $\{u_\nu^{(\sigma,j)}: |\operatorname
{Im}\lambda|\leq\gamma\}$. It is known (see \cite{3}) that the
total algebraic multiplicity of all the eigenvalues of the pencil
$\mathfrak A^r$ in the strip $\{\lambda\in \Bbb C:|
\operatorname{Im} \lambda|\leq\gamma\ \}$ is even for any
$\gamma\geq 0$; we denote the multiplicity by $2M^r(\equiv 2
M^r_\gamma)$. There is a basis
\begin{equation}\label{bu}
u_1^+,\dots,u_{M^r}^+, u_1^-,\dots, u_{M^r}^-
\end{equation}
in the space $\mathcal W_\gamma(\Pi^r)$ obeying
\begin{equation}\label{oc}
 q^r(\chi u_j^\pm, \chi u_h^\pm)=\mp i\delta_{j,h},\quad
 q^r(\chi u_j^\pm,\chi u_h^\mp)=0,\quad j,h=1,\dots,M^r,
\end{equation}
(see \cite{3},\cite{11}); here $\chi$ is the cut-off function from
Proposition~\ref{p2.1}. One can consider the cylinder $\Pi^r$ as a
generalized waveguide. The space $\mathcal W_\gamma(\Pi^r)$ is
called the space of waves. The quantity  $iq^r (\chi u, \chi u)$
represents the energy flow transferred by the wave $u\in\mathcal
W_\gamma(\Pi^r)$ through the cross-section $\Omega^r$ of the
cylinder $\Pi^r$. Thus $u_1^+,\dots,u_{M^r}^+$ are incoming waves
and $u_1^-,\dots, u_{M^r}^-$ are outgoing waves for the
problem~(\ref{lp}).

The following proposition is a variant of Proposition 3.1.4 and
Theorem 3.2.1 from \cite{3}.
\begin{prop}\label{p2.2}
 Assume that $\gamma>0$,  the line $\Bbb
R+i\gamma$ is free of the spectrum of the pencil $\mathfrak A^r$,
and $\{F,G\}\in\mathcal R^l_\gamma(\Pi^r)\cap \mathcal
R^l_{-\gamma} (\Pi^r)$.  Then  a solution
 $u\in\mathcal D_{-\gamma}^\ell (\Pi^r)$ to the problem (\ref{lp}) admits the representation
\begin{equation}\label{asu}
u=\sum_{j=1}^{M^r}\{a_j(F,G) u^+_j+b_j(F,G)  u^-_j \} +v,
\end{equation}
where the functions $u^\pm_j$ form a basis in $\mathcal
W_\gamma(\Pi^r)$ and satisfy (\ref{oc}),  $v$ is a solution to the
 problem (\ref{lp}) in  $\mathcal D_\gamma^l(\Pi^r)$. The functionals $a_1,\dots, a_{M^r}$ and
 $b_1,\dots,b_{M^r}$ are continuous on $\mathcal R^l_\gamma(\Pi^r)\cap \mathcal
R^l_{-\gamma} (\Pi^r)$ and
\begin{equation}\label{c1}
\begin{split}
a_j (F,G) &=i(F,u^+_j)_{\Pi^r}+i(G,Q^r u^+_j)_{\partial\Pi^r},\\
b_j(F,G)&=-i (F,u^-_j)_{\Pi^r}-i(G,Q^r u^-_j)_{\partial\Pi^r},
\end{split}
\end{equation}
where $Q^r$ is the same as in the Green formula (\ref{GF}).
\end{prop}

\section{The structure of solutions to the problem (\ref{1})}
\subsection{Construction of a model problem in $\Pi^r$}\label{sGFg}
 In this subsection we construct a differential operator $\{\mathfrak
L^r_T,\mathfrak B^r_T\}$ in $\Pi^r$ such that the following
conditions are satisfied: (i) $\{\mathfrak L^r_T,\mathfrak
B^r_T\}$ coincides with $\{\mathcal L,\mathcal B\}$ on the set
$\{(y,t)\in\overline \Pi^r: t>T+3\}$; (ii) $\{\mathfrak
L^r_T,\mathfrak B^r_T\}$ coincides with $\{L^r,B^r\}$ on the set
$\{(y,t)\in\overline \Pi^r: t<T\}$; (iii) the norm $\|\Delta_T^r;
\mathcal D^l_\gamma(\Pi^r)\to\mathcal R^l_\gamma (\Pi^r) \|$ of
the operator
\begin{equation}\label{Delta}
\Delta_T^r=\{L^r,B^r\}-\{\mathfrak L^r_T,\mathfrak B^r_T\}
\end{equation}
tends to zero as $T\to +\infty$; (iv) for $u,v\in
C_c^\infty(\overline\Pi^r)$ and sufficiently large $T$ the
self-adjoint Green Formula
\begin{equation}\label{GFg}
({\mathfrak  L}^r_T u,v)_{\Pi^r}+({\mathfrak  B}^r_T u,{\mathfrak
Q}^r_T v)_{\partial \Pi^r}=(u,{\mathfrak L}^r_T
v)_{\Pi^r}+({\mathfrak Q}^r_T u, {\mathfrak B}^r_T v)_{\partial
\Pi^r}
\end{equation}
holds  with some $m\times k$-matrix $\mathfrak Q^r_T$ of
differential operators.

Recall that  $\psi_T(y,t)$ is a cutoff function,
$\psi_T(y,t)\equiv\psi(t-T)$ for $(y,t)\in\overline\Pi^r_+$, where
$\psi\in C^\infty(\Bbb R)$, $\psi(t)=0$ for $t<1$ and $\psi(t)=1$
for $t>2$. Let $\mathfrak L^r_T=L^r-\psi_T(L^r-\mathcal L)\psi_T$,
where the operator $\psi_T\mathcal L\psi_T$ is extended from
$\overline\Pi^r_+$ to the whole cylinder $\overline\Pi^r$ by zero.
First we find a Dirichlet system  $\EuScript P^r_T$ and an
operator $\EuScript T^r_T$ such that
\begin{equation}\label{GFga}
({\mathfrak  L}^r_T u,v)_{\Pi^r}+({\EuScript  P}^r_T u,{\EuScript
T}^r_T v)_{\partial \Pi^r}=(u,{\mathfrak L}^r_T
v)_{\Pi^r}+({\EuScript T}^r_T u, {\EuScript P}^r_T v)_{\partial
\Pi^r}
\end{equation}
for $u,v\in C_c^\infty(\overline\Pi^r)$. Then we compose
$\mathfrak B^r_T$ and $\mathfrak Q^r_T$ from the rows of
$\EuScript P^r_T$ and $\EuScript T^r_T$ (by analogy with
(\ref{3}), (\ref{4} )) and derive (\ref{GFg}) from (\ref{GFga}).

Denote $ \EuScript N^r_T=N^r-\psi_T(N^r-\mathcal N)\psi_T$ and $
 \EuScript D^r_T=\mathcal D^r-\psi_T(\mathcal D^r-\mathcal
 D)\Psi_T$
(the operators $\psi_T\mathcal N\psi_T$ and $\psi_T\mathcal
D\psi_T$ are extended to $\overline\Pi^r$ by zero). It is clear
that $\EuScript D^r_T\equiv\mathcal D^r$ and $\mathcal D^r_T$ is
the Dirichlet system on $\partial\Pi^r$. Since $\mathcal D^r$
consists of normal derivatives, we have $[\mathcal D^r, \psi_T]=0$
on $\partial \Pi^r$; here $[a,b]=ab-ba$. Thus, substituting
$\psi_T u$ and $\psi_T v$ for $u$ and $v$ in (\ref{i.2}), we
obtain
\begin{equation}\label{aa}
\begin{split}
(\psi_T\mathcal L\psi_T u, v)_{\Pi^r}+&(\psi_T\mathcal N\psi_T
u,\mathcal D^r v)_{\partial \Pi^r}\\&=(u,\psi_T\mathcal L\psi_T
v)_{\Pi^r}+(\mathcal D^r u,\psi_T\mathcal N\psi_T  v)_{\partial
\Pi^r}.
\end{split}
\end{equation}
By the same arguments from
\begin{equation}\label{bb}
( L^r u, v)_{\Pi^r}+( N^r u,\mathcal D^r v)_{\partial \Pi^r}=(u,
L^r v)_{\Pi^r}+(\mathcal D^r u, N^r  v)_{\partial \Pi^r}
\end{equation}
 we get
\begin{equation}\label{cc}
\begin{split}
(\psi_T L^r\psi_T u, v)_{\Pi^r}+&(\psi_T N^r\psi_T u,\mathcal D^r
v)_{\partial \Pi^r}\\&=(u,\psi_T L^r\psi_T v)_{\Pi^r}+(\mathcal
D^r u,\psi_T N^r \psi_T  v)_{\partial \Pi^r}.
\end{split}
\end{equation}
Adding (\ref{aa}) and (\ref{bb}) and subtracting (\ref{cc}), we
arrive at the formula
\begin{equation}\label{dd}
(\mathfrak L^r_T u, v)_{\Pi^r}+(\EuScript N^r_T u,\mathcal D^r_T
v)_{\partial \Pi^r}=(u, \mathfrak L^r v)_{\Pi^r}+(\mathcal D^r_T
u, \EuScript N^r_T v)_{\partial \Pi^r}.
\end{equation}
Recall that $\mathcal P=\mathcal R\mathcal D$ and $P^r=R^r
\mathcal D^r$. For sufficiently large $T$ we put $\EuScript
R^r_T=R^r+\psi_T(\mathcal R-R^r)\psi_T$. Due to the stabilization
of $\mathcal R$ in $\Pi^r_+$, the matrix $\EuScript P^r_T\equiv
\EuScript R^r_T\mathcal D^r$ is a Dirichlet system on $\partial
\Pi^r$, and there exists a differential operator $(\EuScript
R^r_T)^{-1}$ such that $(\EuScript R^r_T)^{-1}\EuScript
R^r_T=\EuScript R^r_T(\EuScript R^r_T)^{-1}=I$; see
Remark~\ref{Rr}. Let $\EuScript T^r_T=(\EuScript
R^r_T)^{-1}_*\EuScript N^r_T$. We have
\begin{equation*}\label{ee}
\begin{split}
(\EuScript N^r_T u,\mathcal D^r_T v)_{\partial \Pi^r}-(\mathcal
D^r_T u, \EuScript N^r_T v)_{\partial \Pi^r} &\\
=((\EuScript R^r_T)^{-1}_*\EuScript N^r_T u,\EuScript
R^r_T\mathcal D^r_T v)_{\partial \Pi^r}&-(\EuScript R^r_T\mathcal
D^r_T u, (\EuScript R^r_T)^{-1}_*\EuScript N^r_T v)_{\partial
\Pi^r}\\
&=(\EuScript T^r_T u,\EuScript P^r_T v)_{\partial
\Pi^r}-(\EuScript P^r_T u, \EuScript T^r_T v)_{\partial \Pi^r}.
\end{split}
\end{equation*}
Together with (\ref{dd}) this implies (\ref{GFga}). Composing the
matrices $\mathfrak B^r_T$ and $\mathfrak Q^r_T$ from the rows of
$\EuScript P^r_T$ and $\EuScript T^r_T$  by the same rule as in
(\ref{3}) and (\ref{4}), we obtain the Green formula (\ref{GFg}).

By the construction of $\{\mathfrak L^r_T,\mathfrak B^r_T\}$ the
conditions (i), (ii), and (iv) given in the beginning of this
subsection are satisfied. Due to the stabilization of $\mathcal
L(x,D_x)$ and $\mathcal R(x,D_x)$ in $\Pi^r_+$  we have
\begin{equation}\label{lim}
 \lim_{T\to+\infty}\|\Delta_T^r;
\mathcal D^l_\gamma(\Pi^r)\to\mathcal R^l_\gamma (\Pi^r) \|=0
\end{equation}
for all $\gamma\in \Bbb R$, condition (iii) is fulfilled.

In the cylinder $\Pi^r$ we consider the model problem
\begin{equation}\label{pp}
\begin{split}
\mathfrak L^r_T(y,t,D_y,D_t) u(y,t)&=\mathfrak F(y,t), \quad
(y,t)\in\Pi^r,\\
 \mathfrak B^r_T (y,t,D_y,D_t) u(y,t)&=\mathfrak G (y,t),\quad
 (y,t)\in\partial \Pi^r.
 \end{split}
\end{equation}
\subsection{The structure of solutions to the model problem~(\ref{pp})}
  Taking into account (\ref{lim}) and the
invertibility of the limit operator (\ref{lo}) (see Proposition
\ref{p2.1}), we get the following assertion.
\begin{prop}\label{p3.1}
Let the operators $\mathcal L(x,D_x)$ and $\mathcal R(x,D_x)$
stabilize in $\Pi^r_+$ and let the line $\Bbb R+ i\gamma$ contain
no eigenvalues of the pencil $\mathfrak A^r$. Assume that $T$ is
sufficiently large. Then the operator
$$
\{\mathfrak L^r_T,\mathfrak B^r_T\}:\mathcal
D_\gamma^\ell(\Pi^r)\to\mathcal R_\gamma^\ell(\Pi^r)
$$
of the problem (\ref{pp}) implements an isomorphism.
\end{prop}
We now introduce functions $z_j^\pm$, which play the same role for
the problem~(\ref{pp})  as the waves $u_j^\pm$
 play for the limit problem~(\ref{lp}).
 Suppose that the assumptions of
Proposition~\ref{p3.1} are fulfilled.
 We set
\begin{equation}\label{3.5}
z_j^\pm=u_j^\pm+\sum_{q=1}^\infty
((A^r_{-\gamma})^{-1}\Delta_T^r)^q u_j^\pm,\quad j=1,\dots,M^r,
\end{equation}
where the waves $\{u^\pm_j:j=1,\dots,M^r\}$ form a basis in
$\mathcal W_\gamma(\Pi^r)$ obeying (\ref{oc}). (Recall that $2
M^r(\equiv 2M^r (\gamma))$ is the total algebraic  multiplicity of
all the eigenvalues of the pencil $\mathfrak A^r$ in the strip
$\{\lambda\in \Bbb C: |\operatorname{Im}\lambda|\leq \gamma\}$.)

Let us discuss the equality~(\ref{3.5}). Note that $\psi_{T-2}
u_j^\pm\in\mathcal D^l_{-\gamma}(\Pi^r)$ and $\Delta_T^r
u_j^\pm=\Delta_T^r\psi_{T-2}u_j^\pm$ with the same cutoff function
$\psi_T$ as in the previous section. By
 virtue of (\ref{lim}) the norm of operator $
 (A^r_{-\gamma})^{-1}\Delta_T^r:
\mathcal D^l_{-\gamma}(\Pi^r)\to\mathcal D^l_{-\gamma}(\Pi^r) $ is
small;  $(A^r_{-\gamma})^{-1}$ is bounded because the spectrum of
$\mathfrak A^r$ is symmetric about the real axis, see
Proposition~\ref{p2.1}, (ii). The series $\sum_{q=1}^\infty
((A^r_{-\gamma})^{-1}\Delta_T^r)^q  u_j^\pm$ converges in the norm
of $\mathcal D^l_{-\gamma}(\Pi^r)$. Consequently,
\begin{equation}\label{mod}
z_j^\pm= u_j^\pm\mod \mathcal D^l_{-\gamma}(\Pi^r).
\end{equation}

\begin{prop}\label{p3.2} Let the assumptions of Proposition~\ref{p3.1} be fulfilled. Then the functions
\begin{equation}\label{zpm}
z_1^+,\dots,z_{M^r}^+,z_1^-,\dots,z_{M^r}^-
\end{equation}
 defined by (\ref{3.5})  are linearly
independent  modulo~$\mathcal D_{-\gamma}^l (\Pi^r)$ solutions to
the homogeneous problem~(\ref{pp}).
\end{prop}
\noindent {\bf Proof.} It is easy to see that the functions
$u_\nu^{(\sigma,j)}$ forming the linear span $\mathcal
W_\gamma(\Pi^r)$ are linearly independent modulo~$\mathcal
D_{-\gamma}^l (\Pi^r)$; see (\ref{2.11}). Thus the elements of the
basis $u_1^+,\dots,u_{M^r}^+,u_1^-,\dots,u_{M^r}^-$ in $\mathcal
W_\gamma(\Pi^r)$ are linearly independent modulo~$\mathcal
D_{-\gamma}^l (\Pi^r)$. Together with the relations~(\ref{mod})
this implies the linear independence of $z_j^\pm$, $j=1,\dots,
M^r$,  modulo~$\mathcal D_{-\gamma}^l (\Pi^r)$.

 Let us show that
$z_j^\pm$ satisfy the homogeneous problem~(\ref{pp}). Consider the
equation
\begin{equation}\label{3.2}
\{\mathfrak L^r_T,\mathfrak B^r_T\}w=\Delta^r_T u_j^\pm.
\end{equation}
The inclusion $\Delta^r_T u_j^\pm\in\mathcal R^l_{-\gamma}(\Pi^r)$
holds for $j=1,\dots, M^r$. The line $\Bbb R-i\gamma$ is free of
the spectrum of $\mathfrak A^r$ (because the spectrum of
$\mathfrak A^r$ is symmetric about the real line). By
Proposition~\ref{p3.1} there exists a unique solution
$w\in\mathcal D_{-\gamma}^l(\Pi^r)$ to the problem (\ref{3.2}).
Multiplying (\ref{3.2}) from left by $(A^r_{-\gamma})^{-1}$ and
using (\ref{Delta}), we get
$$
(I-(A^r_{-\gamma})^{-1}\Delta^r_T)w=(A^r_{-\gamma})^{-1}\Delta_T^r
u_j^\pm.
$$
 Thanks to (\ref{lim}) the operator $I-(A^r_{-\gamma})^{-1}\Delta^r_T$
is invertible. We  write $(I-(A^r_{-\gamma})^{-1}\Delta^r_T)^{-1}$
as the Neumann series and obtain
$$
w=\bigl(I-(A^r_{-\gamma})^{-1}\Delta^r_T\bigr)^{-1}(A^r_{-\gamma})^{-1}\Delta_T^r
u_j^\pm=\sum_{q=1}^\infty((A^r_{-\gamma})^{-1}\Delta_T^r)^q
u_j^\pm.
$$
Keeping in mind that $\{L^r,B^r\}u_j^\pm=0$ and (\ref{Delta}), we
deduce from (\ref{3.2}) that $\{\mathfrak L^r_T,\mathfrak
B^r_T\}(w+u_j^\pm)=0$. It remains to note that $z_j^\pm\equiv
w+u_j^\pm$.
 \qed

 Introduce the form
\begin{equation}\label{pr}
\begin{split}
p^r_T(u,v)=({\mathfrak  L}^r_T u,&v)_{\Pi^r}+({\mathfrak  B}^r_T
u,{\mathfrak Q}^r_T v)_{\partial \Pi^r}\\&-(u,{\mathfrak L}^r_T
v)_{\Pi^r}-({\mathfrak Q}^r_T u, {\mathfrak B}^r_T v)_{\partial
\Pi^r}.
\end{split}
\end{equation}
It is easy to see that  $p^r_T(u,v)=0$  for $u\in\mathcal
D^l_{-\gamma}(\Pi^r)$ and $v\in\mathcal D^l_{\gamma}(\Pi^r)$
 (indeed, for such functions the Green formula
(\ref{GFg}) is fulfilled).

 \begin{prop}\label{p3.3} Let the assumptions of
Proposition~\ref{p3.1} be fulfilled. Then the functions
(\ref{zpm}) satisfy the conditions
\begin{equation}
\label{ocz} p^r_T(\chi z_h^\pm,\chi z_j^\pm)=\mp
i\delta_{h,j},\quad p^r_T(\chi z_h^\pm,\chi z_j^\mp) =0,\quad
h,j=1,\dots,M^r,
\end{equation}
where $\chi \in C^\infty(\Bbb R)$, $\chi(t)=1$ for $t\geq 2$ and
$\chi(t)=0$ for $t\leq 1$.The equalities~(\ref{ocz}) do not depend
on the choice of $\chi$.
 \end{prop}

\noindent{\bf Proof.} Since the waves $u^\pm_j$
  satisfy the homogeneous problem (\ref{lp}), we have
$q^r(u^\pm_h,u^\pm_j)=0$ and
\begin{equation}\label{3.3}
-q^r(\chi u^\pm_h,u^\pm_j)=q^r((1-\chi)u^\pm_h,u^\pm_j),
\end{equation}
where $q^r$ is from (\ref{2.12}). Note that operator $\{L^r,B^r\}$
coincide with  $\{\mathfrak L^r_T,\mathfrak B^r_T\}$ on the
support of $(1-\chi)u^\pm_h$. This allows us to write (\ref{3.3})
in the form
\begin{equation}\label{+1}
-q^r(\chi u^\pm_h,u^\pm_j)=p_T^r((1-\chi)u^\pm_h,u^\pm_j).
\end{equation}
Due to (\ref{mod}) and $\chi u^\pm_h\in\mathcal
D^l_{\gamma}(\Pi^r)$ the Green formula (\ref{GFg}) is valid on the
pairs
\begin{eqnarray*}
& &\{u,v\}=\{(1-\chi)(z^\pm_h-u^\pm_h),z^\pm_j\},\\
& &\{u,v\}=\{(1-\chi)(z^\pm_h-u^\pm_h), z^\pm_j-u^\pm_j\},\\
& &\{u,v\}=\{(1-\chi)z^\pm_h, z^\pm_j-u^\pm_j\}.
\end{eqnarray*}
 Thus $p^r_T(u,v)=0$ on the same pairs, and
\begin{equation}\label{+2}
p_T^r((1-\chi)u^\pm_h,u^\pm_j)=p_T^r((1-\chi) z^\pm_h,z^\pm_j).
\end{equation}
 Thanks to Proposition~\ref{p3.2} we have
$p^r_T(z^\pm_h,z^\pm_j)=0$. Therefore, from (\ref{+1}) and
(\ref{+2}) we get
$$
-q^r(\chi u^\pm_h,u^\pm_j)=-p_T^r(\chi z^\pm_h,z^\pm_j).
$$
Finally we obtain
$$
q^r(\chi u^\pm_h,\chi u^\pm_j)=q^r(\chi u^\pm_h,
u^\pm_j)=p_T^r(\chi z^\pm_h,z^\pm_j)=p_T^r(\chi z^\pm_h,\chi
z^\pm_j).
$$
To establish the first equality in (\ref{ocz}) it remains to use
(\ref{oc}). In a similar way one can prove the second equality in
(\ref{ocz}). \qed

The first assertion of the following theorem is a variant of
Theorem 6.2 from \cite{1}; see also \cite[Theorem 8.5.7]{2}.
 \begin{thm}\label{mp}
Assume that the operators $\mathcal L$ and $\mathcal R$ stabilize
in $\Pi^r$. Let $\gamma>0$ and let the line $\Bbb R+i\gamma$ be
free of the spectrum of the pencil $\mathfrak A^r$.  Then for
sufficiently large $T$ the following assertions hold.

\noindent{\rm(i)} A solution~$u\in\mathcal D_{-\gamma}^l (\Pi^r)$
to the problem (\ref{pp}) with right-hand side $\{\mathfrak
F,\mathfrak G\}\in \mathcal R^l_\gamma(\Pi^r)\cap \mathcal
R^l_{-\gamma} (\Pi^r)$ admits the representation
\begin{equation}\label{asz}
u=\sum_{j=1}^{M^r} \{a_j(\mathfrak F,\mathfrak
G)\,z_j^++b_j(\mathfrak F,\mathfrak G)\,z_j^-\}+v,
\end{equation}
where $v$ is a solution to the same problem in $\mathcal
D_\gamma^l(\Pi^r)$, the waves $z_j^\pm$ are defined by
(\ref{3.5}), and $2 M^r$ is the total algebraic multiplicity of
the eigenvalues of the pencil $\mathfrak A^r$ in the strip
$\{\lambda\in \Bbb C:|\operatorname{Im}|<\gamma\}$.

\noindent{\rm(ii)} The functionals $a_1,\dots,a_{M^r}$ and
$b_1,\dots,b_{M^r}$ are continuous  on $\mathcal
R^l_\gamma(\Pi^r)\cap \mathcal R^l_{-\gamma} (\Pi^r)$ and
\begin{equation}\label{d1}
\begin{split}
a_j (\mathfrak F,\mathfrak G) &=i(\mathfrak F,z^+_j)_{\Pi^r}+i(\mathfrak G,\mathfrak Q^r_T z^+_j)_{\partial\Pi^r},\\
b_j(\mathfrak F,\mathfrak G)&=-i (\mathfrak
F,z^-_j)_{\Pi^r}-i(\mathfrak G,\mathfrak Q^r_T
z^-_j)_{\partial\Pi^r},
\end{split}
\end{equation}
where $\mathfrak Q^r_T$ is the same as in the Green formula
(\ref{GFg}).
 \end{thm}

\noindent{\bf Proof.} (i) Since the spectrum of $\mathfrak A^r$ is
symmetric about the real line, the conditions of theorem guaranty
that the line $\Bbb R- i\gamma$ is free of the spectrum. From the
second assertion of Proposition~\ref{p2.1} and (\ref{lim}) it
follows that
\begin{eqnarray*}\
& &\|(A_\gamma^r)^{-1}\Delta_T^r;\mathcal
D_\gamma^\ell(\Pi^r)\to\mathcal D_\gamma^\ell(\Pi^r)\|<1,\\
& &\|(A_{-\gamma}^r)^{-1}\Delta_T^r;\mathcal
D_{-\gamma}^\ell(\Pi^r)\to\mathcal D_{-\gamma}^\ell(\Pi^r)\|<1.
\end{eqnarray*}
Solutions $u\in\mathcal D_{-\gamma}^\ell (\Pi^r)$ and
$v\in\mathcal D_\gamma^\ell (\Pi^r)$ to the problem~(\ref{pp})
satisfy the equations
\begin{equation*}
u =(A^r_{-\gamma})^{-1}\Delta^r_T
u+(A^r_{-\gamma})^{-1}\{\mathfrak F,\mathfrak G\},\quad v
=(A^r_\gamma)^{-1}\Delta^r_T v+(A^r_\gamma)^{-1}\{\mathfrak
F,\mathfrak G\},
\end{equation*}
where $\Delta^r_T$ is from (\ref{Delta}). Let us solve this
equations by the method of successive approximations. We set
\begin{equation*}
\begin{split}
 u_{n+1}&=(A^r_{-\gamma})^{-1}\Delta^r_T u_n+u_0,\quad
u_0=(A^r_{-\gamma})^{-1}\{\mathfrak F,\mathfrak
G\},\\
v_{n+1} &=(A^r_\gamma)^{-1}\Delta^r_T v_n+v_0,\quad
v_0=(A^r_\gamma)^{-1}\{\mathfrak F,\mathfrak G\}.
\end{split}
\end{equation*}
By Proposition~\ref{p2.2} we have
\begin{equation}\label{3.13}
u_0=i \sum_{j=1}^{M^r} \{a_j(\mathfrak F,\mathfrak G) u_j^+ +b_j
(\mathfrak F,\mathfrak G)u_j^-\}+v_0.
\end{equation}
Let us write the formulas~(\ref{c1}) for $a_j(\mathfrak
F,\mathfrak G)$ and $b_j(\mathfrak F,\mathfrak G)$ in the form
\begin{equation*}
a_j(\mathfrak F,\mathfrak G)=i q^r(v_0,u^+_j), \quad b_j(\mathfrak
F,\mathfrak G)=-i q^r(v_0,u^-_j),
\end{equation*}
where $q^r$ is from (\ref{2.12}). We prove by induction that
\begin{equation}\label{3.14}\begin{split}
u_n=v_n+i \sum_{j=1}^{M^r} \sum_{m=0}^n
((A^r_{-\gamma})^{-1}\Delta_T^r)^m\{q^r(v_{n-m},& u^+_j)u_j^+ \\&-
q^r(v_{n-m}, u^-_j)u_j^-\}.
\end{split}
\end{equation}
If $n=0$ then (\ref{3.14}) coincides with~(\ref{3.13}). We suppose
that~(\ref{3.14}) holds for $n$ and show that it remains valid for
$n+1$. From~(\ref{3.14}) we get
\begin{equation}\label{3.15}
\begin{split}
(&A^r_{-\gamma})^{-1}\Delta_T^r u_n=
(A^r_{-\gamma})^{-1}\Delta_T^r v_n\\&+i\sum_{j=1}^{M^r}
\sum_{m=0}^n ((A^r_{-\gamma})^{-1}\Delta_T^r)^{m+1}\{q^r(v_{n-m},
u^+_j)u_j^+- q^r(v_{n-m}, u^-_j)u_j^-\}.
\end{split}
\end{equation}
Using Proposition~\ref{p2.2}, we
represent~$(A^r_{-\gamma})^{-1}\Delta_T^r v_n$ in the form
\begin{equation}\label{3.16}\begin{split}
(&A^r_{-\gamma})^{-1}\Delta_T^r v_n=(A_\gamma^r)^{-1}\Delta^r_T
v_n\\&+i\sum_{j=1}^{M^r}
\bigl\{q^r\bigl((A^r_{-\gamma})^{-1}\Delta_T^r v_{n},
u^+_j\bigr)u_j^+- q^r\bigl((A^r_{\gamma})^{-1}\Delta_T^r v_{n},
u^-_j\bigr)u_j^-\bigr\}.
\end{split}
\end{equation}
 Taking into account~(\ref{3.13}),  (\ref{3.16}) and the formulas
for $u_{n+1}$ and $v_{n+1}$, we pass from~(\ref{3.15}) to the
equality~(\ref{3.14}) with  $n$ replaced by $n+1$. The
formula~(\ref{3.14}) is proved.

Substituting
$v_{n-m}=\sum_{h=0}^{n-m}((A^r_{\gamma})^{-1}\Delta_T^r)^h v_0$
into (\ref{3.14}) we obtain
\begin{equation}\label{3.14+}\begin{split}
u_n=v_n+i \sum_{j=1}^{M^r} \sum_{m=0}^n
((A^r_{-\gamma})^{-1}&\Delta_T^r)^m\Bigl\{q^r\bigl(\sum_{h=0}^{n-m}((A^r_{\gamma})^{-1}\Delta_T^r)^h
v_0, u^+_j\bigr)u_j^+ \\&-
q^r\bigl(\sum_{h=0}^{n-m}((A^r_{\gamma})^{-1}\Delta_T^r)^h v_0,
u^-_j\bigr)u_j^-\Bigr\}.
\end{split}
\end{equation}
The series $ \sum_{h=0}^\infty ((A^r_{\gamma})^{-1}\Delta_T^r)^h
v_0 $ converges in the norm of $\mathcal D_\gamma^\ell(\Pi^r)$,
moreover,
$\{L^r,B^r\}\sum_{h=0}^\infty((A^r_{\gamma})^{-1}\Delta_T^r)^h
v_0\in \mathcal R_\gamma^\ell(\Pi^r)\cap\mathcal
R_{-\gamma}^\ell(\Pi^r)$. Using the argument given
after~(\ref{3.5}), we justify the passage to the limit
in~(\ref{3.14+}) as $n\to \infty$. As a result we get the
representation~(\ref{asz}), where the functions $u\in\mathcal
D_{-\gamma}^\ell(\Pi^r)$ and $v\in\mathcal D_\gamma^\ell(\Pi^r)$
satisfy the problem~(\ref{pp}), and
\begin{eqnarray*}
a_j(\mathfrak F,\mathfrak G)& = &i q^r\bigl(\sum_{h=0}^\infty
((A^r_{\gamma})^{-1}\Delta_T^r)^h v_0, u_j^+), \\b_j(\mathfrak
F,\mathfrak G)&= &-iq^r\bigl(\sum_{h=0}^\infty
((A^r_{\gamma})^{-1}\Delta_T^r)^h v_0, u_j^-).
\end{eqnarray*}
The assertion (i) is proved.

Let us establish the formulas (\ref{d1}). Due to
Proposition~\ref{p3.2} we have
\begin{equation}\label{*}
(\mathfrak F, z_j^\pm)_{\Pi^r}+(\mathfrak G,\mathfrak Q^r_T
z_j^\pm)_{\partial\Pi^r}=p^r_T(u,z_j^\pm),
\end{equation}
where $u$ is the same as in (\ref{asz}). Let $\chi \in
C^\infty(\Bbb R)$, $\chi(t)=1$ for $t\geq 2$ and $\chi(t)=0$ for
$t\leq 1$. From (\ref{mod}) and $(1-\chi)u_j^\pm \in
D^l_{\gamma}(\Pi^r)$ we obtain $(1-\chi)z_j^\pm\in
D^l_{\gamma}(\Pi^r)$. Then the inclusion $u\in \mathcal
D^l_{\gamma}(\Pi^r)$ implies $p^r_T(u,(1-\chi)z_j^\pm)=0$.
Together with (\ref{asz}) this allows us to write (\ref{*}) in the
form
\begin{equation*}
\begin{split}
(\mathfrak F,& z_j^\pm)_{\Pi^r}+(\mathfrak G,\mathfrak Q^r_T
z_j^\pm)_{\partial\Pi^r}\\
&= p^r_T(\sum_{h=1}^{M^r} \{a_h(\mathfrak F,\mathfrak
G)\,z_h^++b_h(\mathfrak F,\mathfrak G)\,z_h^-\}+v,\chi z_j^\pm ).
\end{split}
\end{equation*}
Note that $p^r_T(v,\chi z_j^\pm)=0$ as far as  $v\in\mathcal
D_{\gamma}^l(\Pi^r)$ and $\chi z_j^\pm\in \mathcal
D_{-\gamma}^l(\Pi^r)$; see~(\ref{mod}). Finally we have
\begin{equation*}
\begin{split}
(\mathfrak F,& z_j^\pm)_{\Pi^r}+(\mathfrak G,\mathfrak Q^r_T
z_j^\pm)_{\partial\Pi^r}\\
&= -p^r_T(\sum_{h=1}^{M^r} \{a_h(\mathfrak F,\mathfrak
G)\,z_h^++b_h(\mathfrak F,\mathfrak G)\,z_h^-\},\chi z_j^\pm )\\&=
-p^r_T(\sum_{h=1}^{M^r} \{a_h(\mathfrak F,\mathfrak G)\,\chi
z_h^++b_h(\mathfrak F,\mathfrak G)\,\chi z_h^-\},\chi z_j^\pm).
\end{split}
\end{equation*}
By applying Proposition~\ref{p3.3}, we complete the proof. \qed

Theorem~\ref{mp} does not allow us to write a structure of $u\in
\mathcal D_\beta^l(\Pi^r)$ with a remainder $v\in \mathcal
D_\gamma^l(\Pi^r)$ if $\beta\ne -\gamma$. Further we correct this
trouble.

Let $\alpha_\nu$, $\nu\in\Bbb Z$, be numbers such that every strip
$\alpha_\nu\leq\operatorname{Im}\lambda<\operatorname{Im}\lambda_\nu$
is free of the spectrum of $\mathfrak A^r$. For sufficiently large
$T$ we set
\begin{equation}\label{w}
w^{(\sigma,j)}_\nu=u^{(\sigma,j)}_\nu+\sum_{q=1}^\infty
((A^r_{\alpha_\nu})^{-1}\Delta^r_T)^q u^{(\sigma,j)}_\nu,
\end{equation}
where the functions $u^{(\sigma,j)}_\nu$ are given in (\ref{2.11})
and satisfy the conditions (\ref{2.13})--(\ref{2.15}). Repeating
the arguments form the proof of Proposition~\ref{p3.2} one can
show that $w^{(\sigma,j)}_\nu$ solves the homogenous model problem
(\ref{pp}).  The functions $w^{(\sigma,j)}_\nu$ do not depend on
the choice of $\alpha_\nu$; indeed,
$(A^r_{\alpha})^{-1}\{F,G\}=(A^r_{\beta})^{-1}\{F,G\}$ provided
that the strip $\alpha\leq\operatorname{Im}\lambda\leq \beta$ is
free of the spectrum of the pencil $\mathfrak A^r$ and
$\{F,G\}\in\mathcal R^l_\alpha (\Pi^r) \cap \mathcal R^l_\beta
(\Pi^r)$  (see e.g. \cite[Proposition 3.1.4]{3}). From the
relations $w^{(\sigma,j)}_\nu = u^{(\sigma,j)}_\nu \mod \mathcal
D_{\alpha_\nu}^l(\Pi^r)$ and the formulas (\ref{2.11}) it follows
the linear independence of functions $w^{(\sigma,j)}_\nu$.

\begin{lm}\label{pwz} Let the assumptions of Theorem~\ref{mp} be fulfilled and let $\lambda_{-M},\dots,\lambda_M$ be all
eigenvalues of $\mathfrak A^r$ from the strip
$-\gamma<\operatorname{Im}\lambda<\gamma$. Then for sufficiently
large $T$ the relations
\begin{equation}\label{wz}
w^{(\tau,p)}_\mu=\sum_{\nu=1}^M\{a^{(\tau,p)}_{\mu,\nu} z^+_\nu
+b^{(\tau,p)}_{\mu,\nu} z^-_\nu\},\\
\end{equation}
hold with the coefficients
\begin{equation}\label{ab}
a_{\mu , \nu}^{(\tau, p)}=i p^r_T(\chi w^{(\tau,p)}_\mu,\chi
z_{\nu}^+),\qquad b_{\mu , \nu}^{(\tau, p)}=-i p^r_T(\chi
w^{(\tau,p)}_\mu,\chi z_{\nu}^-),
\end{equation}
where $\mu=-M,\dots,M$, $p=1,\dots,J_\mu$, and
$\tau=0,\dots,\varkappa_{p\mu}-1$;  $\chi\in C^\infty (\Bbb R)$,
$\chi(t)=1$ for $t>2$ and $\chi(t)=0$ for $t<1$.
\end{lm}

 \noindent {\bf Proof.} Since $w^{(\tau,p)}_\mu\in
\mathcal D_{\alpha_\mu}^l(\Pi^r)$, where $\gamma>\alpha_\mu\geq
-\gamma$, we have $\chi w^{(\tau,p)}_\mu\in \mathcal
D_{-\gamma}^l(\Pi^r)$ and $(\chi-1)w^{(\tau,p)}_\mu\in \mathcal
D_{\gamma}^l(\Pi^r)$. We put $\{\mathfrak F,\mathfrak G\}
=-\{\mathfrak L^r_T,\mathfrak B^r_T\}\chi w^{(\tau,p)}_\mu$. It is
clear that $\{\mathfrak F,\mathfrak G\}\in\mathcal
R_\gamma^l(\Pi^r)\cap\mathcal R_{-\gamma}^l(\Pi^r)$ and
$\{\mathfrak L^r_T,\mathfrak B^r_T\}(1-\chi)
w^{(\tau,p)}_\mu=\{\mathfrak F,\mathfrak G\}$. By
Proposition~\ref{p3.1} and Theorem~\ref{mp} we have
$$
(1-\chi) w^{(\tau,p)}_\mu=\sum_{\nu=1}^M\{a^{(\tau,p)}_{\mu,\nu}
z^+_\nu +b^{(\tau,p)}_{\mu,\nu} z^-_\nu\}-\chi w^{(\tau,p)}_\mu.
$$
This leads to (\ref{wz}). The equalities (\ref{ab}) are readily
apparent from (\ref{wz}) and  Proposition~\ref{p3.3}.
  \qed
\begin{prop}\label{bang-up}
 Let $\chi \in C^\infty(\mathbb R)$, $\chi (t)=1$ for
$t\geq2$ and $\chi (t)=0$ for $t\leq 1$. The functions
$w_\nu^{(\sigma,\nu)}$ given in (\ref{w}) satisfy the following
conditions:
\begin{eqnarray}
\label{w.1}p_T^r(\chi w_\nu^{(\sigma,j)},\chi w_\mu^{(\tau,p)})  =i\delta_{-\nu,\mu}\delta_{j,p}\delta_{\varkappa_{j\nu}-1-\sigma,\tau},\quad |\nu|>\nu_0, |\mu|>\nu_0,\\
\label{w.2} p_T^r(\chi w_\nu^{(\sigma,j)},\chi
w_\mu^{(\tau,p)})=\pm
i\delta_{\nu,\mu}\delta_{j,p}\delta_{\varkappa_{j\nu}-1-\sigma,\tau},\quad |\nu|\leq \nu_0,|\mu|\leq \nu_0,\\
\label{w.3}p_T^r(\chi w_\nu^{(\sigma,j)},\chi w_\mu^{(\tau,p)})
=0,\quad |\nu|\leq \nu_0,|\mu|>\nu_0.
\end{eqnarray}
In (\ref{w.2}) the sign depends on $\nu$ and $j$ and coincides
with the sign in (\ref{2.14}). The conditions~(\ref{w.1}) --
(\ref{w.3}) do not depend on the choice of $\chi$.
\end{prop}
\noindent{\bf Proof.} First we prove that $p_T^r(\chi
w_\nu^{(\sigma,j)},\chi w_\mu^{(\tau,p)})=0$ if $\operatorname
{Im} (\lambda_\nu+\lambda_\mu)\ne 0$.

Let $\operatorname {Im} (\lambda_\nu+\lambda_\mu)> 0$. In this
case one can choose $\alpha_\nu$ and $\alpha_\mu$ (see~(\ref{w}))
 such that
$\alpha_\nu+\alpha_\mu>0$. Then for the functions $u:=\chi
w_\nu^{(\sigma,j)}\in \mathcal D_{\alpha_\nu}^l(\Pi^r)$ and
$v:=\chi w_\mu^{(\tau,p)}\in \mathcal D_{\alpha_\mu}^l(\Pi^r)$ the
Green formula (\ref{GFg}) holds. This implies $p_T^r(\chi
w_\nu^{(\sigma,j)},\chi w_\mu^{(\tau,p)})=0$.

Let us consider the case $\operatorname {Im}
(\lambda_\nu+\lambda_\mu)< 0$. One can choose $\beta_\nu$ and
$\beta_\mu$ such that $\beta_\nu>\operatorname {Im}\lambda_\nu$,
$\beta_\mu>\operatorname {Im}\lambda_\mu$, and
$\beta_\nu+\beta_\mu<0$. Then  $(1-\chi) w_\nu^{(\sigma,j)}\in
\mathcal D_{\beta_\nu}^l(\Pi^r)$,  $(1-\chi) w_\mu^{(\tau,p)}\in
\mathcal D_{\beta_\mu}^l(\Pi^r)$ and for $u:=(1-\chi)
w_\nu^{(\sigma,j)}$ and $v:=(1-\chi) w_\mu^{(\tau,p)}$ the Green
formula (\ref{GFg}) holds. This implies $p_T^r((1-\chi)
w_\nu^{(\sigma,j)},(1-\chi) w_\mu^{(\tau,p)})=0$. Since
  the Green formula
(\ref{GFg}) holds for $u,v\in C_c^\infty (\overline G)$ and
$p^r_T(w_\nu^{(\sigma,j)},w_\mu^{(\tau,p)})=0$, we get
$$
p_T^r((1-\chi) w_\nu^{(\sigma,j)},(1-\chi)
w_\mu^{(\tau,p)})=p_T^r((1-\chi)
w_\nu^{(\sigma,j)},w_\mu^{(\tau,p)})=-p_T^r(\chi
w_\nu^{(\sigma,j)},\chi w_\mu^{(\tau,p)}).
$$
Thus $p_T^r(\chi w_\nu^{(\sigma,j)},\chi w_\mu^{(\tau,p)})=0$ if
$\operatorname {Im} (\lambda_\nu+\lambda_\mu)\ne 0$.

 Let $\operatorname {Im}
(\lambda_\nu+\lambda_\mu)=0$. Without loss of generality we can
assume that $\operatorname{Im}\lambda_\nu\geq 0$. Then
$\alpha_\mu<-\operatorname{Im}\lambda_\nu\leq 0$. We set
$\{F,G\}=\{L^r, B^r\}(1-\chi)w_\nu^{(\sigma,j)}$.
 It is clear that $\{L^r, B^r\}(1-\chi)w_\nu^{(\sigma,j)}=\{\mathfrak
L_T^r,\mathfrak B_T^r\}(1-\chi)w_\nu^{(\sigma,j)}\in \mathcal
R^l_{-\alpha_\mu} (\Pi^r)\cap \mathcal R^l_{\alpha_\mu} (\Pi^r)$.
Write down the asymptotic of the solution
$(1-\chi)w_\nu^{(\sigma,j)}\in \mathcal D_{-\alpha_\mu}^l(\Pi^r)$
to the problem $\{L^r, B^r\}u=\{F,G\}$. We have
$$
 (1-\chi)w_\nu^{(\sigma,j)}=\sum_{h=-M}^M\sum_{s=1}^{J_h}\sum_{\delta=0}^{\varkappa_{sh}-1} c_h^{(\delta,s)}
u_h^{(\delta,s)}\quad \mod\  \mathcal D_{\alpha_\mu}^l(\Pi^r),
$$
where $2M$  stands for the total algebraic multiplicity of all
eigenvalues of $\mathfrak A^r$ in the strip
$-\alpha_\mu>\operatorname{Im}\lambda>\alpha_\mu$; see e.g.
\cite[Proposition 3.1.4]{3}. Note that $c_\nu^{(\sigma,j)}=1$
because of the inclusion $(w^{(\sigma,j)}_\nu -
u^{(\sigma,j)}_\nu) \in \mathcal D_{\alpha_\nu}^l(\Pi^r)$.
Therefore,
\begin{eqnarray*}
-p^r_T(\chi w^{(\sigma,j)}_\nu, \chi w^{(\tau,p)}_\mu)=
p^r_T((1-\chi)w^{(\sigma,j)}_\nu,
(1-\chi)w^{(\tau,p)}_\mu)\\=q^r((1-\chi)w^{(\sigma,j)}_\nu,
(1-\chi)w^{(\tau,p)}_\mu)
=q^r(\sum_{h=-M}^M\sum_{s=1}^{J_h}\sum_{\delta=0}^{\varkappa_{sh}-1}
c_h^{(\delta,s)} u_h^{(\delta,s)}, (1-\chi)u^{(\tau,p)}_\mu)\\
=-q^r(\sum_{h=-M}^M\sum_{s=1}^{J_h}\sum_{\delta=0}^{\varkappa_{sh}-1}
c_h^{(\delta,s)} \chi u_h^{(\delta,s)}, \chi u^{(\tau,p)}_\mu)
\end{eqnarray*}
(in the next-to-last equality we used that $q^r(u,v)=0$ for
$u\in\mathcal D^l_{-\alpha_\mu}(\Pi^r)$ and $v\in\mathcal
D^l_{\alpha_\mu}(\Pi^r)$). Taking into account the equality
$c_\nu^{(\sigma,j)}=1$ and the relations
(\ref{2.13})--(\ref{2.15}),  we complete the proof. \qed

\begin{thm}\label{tasw}
Assume that $\mathcal L$ and $\mathcal R$ stabilize in $\Pi^r_+$.
We also suppose that $\beta>\alpha$ and the lines $\Bbb R+i\alpha$
and $\Bbb R+i\beta$ contain no eigenvalues of the pencil
$\mathfrak A^r$. Let $\lambda_K,\dots,\lambda_M$ be all
eigenvalues of $\mathfrak A^r$ from the strip
$\alpha<\operatorname{Im}\lambda<\beta$ and $\{\mathfrak
F,\mathfrak G\}\in \mathcal R^l_\alpha(\Pi^r)\cap
R^l_\beta(\Pi^r)$. Then  for sufficiently large $T$ the following
assertions hold.

\noindent {\rm (i)} A solution $u\in\mathcal D^l_\alpha(\Pi^r)$ to
the model problem (\ref{pp}) admits the representation
\begin{equation}\label{asw}
u=\sum_{\nu=K}^M\sum_{j=1}^{J_\nu}\sum_{\sigma=0}^{\varkappa_{j\nu}-1}
d_\nu^{(\sigma,j)}(\mathfrak F,\mathfrak G)\,w_\nu^{(\sigma,j)}+v,
\end{equation}
where $v$ is a solution to the same problem in $\mathcal
D^l_\beta(\Pi^r)$.

\noindent {\rm (ii)} The coefficients
$d_\nu^{(\sigma,j)}(\mathfrak F,\mathfrak G)$ in (\ref{asw}) can
be found by the formulas
\begin{equation}\label{wd1}
\begin{split}
d_\nu^{(\sigma,j)}(\mathfrak F,\mathfrak G) =i\{(\mathfrak
F,&w_{-\nu}^{(\varkappa_{j\nu}-\sigma-1,j)})_{\Pi^r}\\&+(\mathfrak
G,\mathfrak Q_T^r
w_{-\nu}^{(\varkappa_{j\nu}-\sigma-1,j)})_{\partial\Pi^r}\},\quad
|\nu|>\nu_0,
\end{split}
\end{equation}
and
\begin{equation}\label{wd2}
\begin{split}
 d_\nu^{(\sigma,j)}(\mathfrak F,\mathfrak G)=\pm i\{
(\mathfrak F,&
w_{\nu}^{(\varkappa_{j\nu}-\sigma-1,j)})_{\Pi^r}\\&+ (\mathfrak
G,\mathfrak Q_T^r
w_{\nu}^{(\varkappa_{j\nu}-\sigma-1,j)})_{\partial\Pi^r}\},\quad
|\nu|\leq\nu_0.
\end{split}
\end{equation}
The sign in (\ref{wd2}) is  the same as in~(\ref{2.14}).
\end{thm}
\noindent{\bf Proof.} Let $\gamma=\max\{-\alpha, \beta\}$. We
again use  the cut-off function $\chi\in C^c_\infty (\overline
G)$, $\chi(t)=1$ for $t\geq 2$ and $\chi(t)=0$ for $t\leq 1$.

 Let us
first consider the case $\gamma=-\alpha$. Let  $v\in\mathcal
D_{\beta}^l(\Pi^r)$ satisfy the model problem~(\ref{pp}). We set
$\{F,G\}:=\{\mathfrak L^r_T,\mathfrak B^r_T\}(1-\chi)v\in \mathcal
R^l_{-\gamma}(\Pi^r)\cap \mathcal R^l_{\gamma}(\Pi^r)$. By
Theorem~\ref{mp} we have
\begin{equation}\label{111}
y=\sum_{j=1}^{-K}\{a_j z^+_j+b_j z^-_j\}+(1-\chi)v,
\end{equation}
where $y\in \mathcal D_{-\gamma}^l (\Pi^r)$.  Due to the linear
independence of $w^{(\sigma,j)}_\nu$ and (\ref{wz}) the waves
$z_\nu^\pm$ can be expressed in terms of $w^{(\sigma,j)}_\nu$.
From (\ref{111}) we get
\begin{equation}\label{112}
u=y+\chi v=\sum_{j=1}^{-K}\{a_j z^+_j+b_j
z^-_j\}+v=\sum_{\nu=K}^{-K}\sum_{j=1}^{J_\nu}\sum_{\sigma=0}^{\varkappa_{j\nu}-1}
d_\nu^{(\sigma,j)}( F, G)\,w_\nu^{(\sigma,j)}+v,
\end{equation}
where $u=(\chi v+ y)\in \mathcal D^l_{\alpha}(\Pi^r)$. Since
$(1-\chi)(u-v)\in\mathcal D^l_{\beta}(\Pi^r)$ and the function
$(1-\chi)w_\nu^{(\sigma,j)}$ is in $\mathcal D^l_{\beta}(\Pi^r)$
only if $ \nu\geq K$, we have $d_\nu^{(\sigma,j)}( F, G)=0$ for
$\nu=M+1,\dots, -K$.  In the case $-\alpha\geq\beta$
the representation (\ref{asw}) is proved.

Consider the case $-\alpha<\beta$. Then $\gamma=\beta$. Applying
Theorem~\ref{mp}, we get the representation
\begin{equation*}
\chi u=\sum_{j=1}^{M}\{a_j z^+_j+b_j z^-_j\}+y, \quad
y\in\mathcal D^l_{\gamma}(\Pi^r),
\end{equation*}
for the solution $\chi u\in \mathcal D^l_{-\gamma}(\Pi^r)$ to the
model problem~(\ref{pp}) with right-hand side $\{F,
G\}:=\{\mathfrak L^r_T,\mathfrak B^r_T\}\chi u\in \mathcal
R^l_{-\gamma}(\Pi^r)\cap \mathcal R^l_{\gamma}(\Pi^r)$. Therefore,
\begin{equation*}
u=\sum_{\nu=-M}^{M}\sum_{j=1}^{J_\nu}\sum_{\sigma=0}^{\varkappa_{j\nu}-1}
d_\nu^{(\sigma,j)}( F, G)\,w_\nu^{(\sigma,j)}+v,
\end{equation*}
where $v=(y+(1-\chi)u)\in  \mathcal D^l_{\beta}(\Pi^r)$. Owing to
the inclusion $\chi (u-v)\in \mathcal D^l_{\alpha}(\Pi^r)$, we
have $d_\nu^{(\sigma,j)}( F, G)=0$ for $\nu=-M,\dots,K-1$. The
 assertion (i) of the theorem is proved.

Using  Proposition~\ref{bang-up} and the
representation~(\ref{asw}) one can prove the formulas (\ref{wd1})
and (\ref{wd2}) for the coefficients; see the proof of (\ref{d1})
in Theorem~\ref{mp}. \qed

\subsection{The structure of solutions to the problem~(\ref{1})}
Let us define the spaces $\mathcal D_\gamma^\ell(G)$ and $\mathcal
R_\gamma^\ell(G)$ by the equalities (\ref{sp}) with $\Pi^r$
replaced by $G$; the space $W_\gamma(G)$ is endowed with the norm
$\|e_\gamma\cdot;H^\ell(G)\|$, where $e_\gamma$ is smooth positive
function in $\overline G$ such that $e_\gamma(y^r,t^r)=\exp\gamma
t^r$ for $(y^r,t^r)\in\bar\Pi^r$.

Assume that the operators  $\mathcal L(x,D_x)$ and $\mathcal
R(x,D_x)$ stabilize in $\Pi_+^1,\dots,\Pi_+^N$. As was shown the
stabilization in $\Pi_+^r$ implies (\ref{lim}). Thus the operator
\begin{equation}\label{o1}
\mathcal A(\gamma)=\{\mathcal L,\mathcal B\}:\mathcal
D_\gamma^\ell(G)\to\mathcal R_\gamma^\ell(G)
\end{equation}
of the problem~(\ref{1}) is continuous.

Proposition~\ref{p3.1} and the well known results of the local
theory of elliptic boundary value problems enable one to prove the
following proposition in the standard way. The proof is omitted.
\begin{prop}\label{pF}
Let the operators $\mathcal L(x,D_x)$ and $\mathcal R(x,D_x)$
stabilize in $\Pi_+^1,\dots,\Pi_+^N$ and let the line $\Bbb R+
i\gamma$ be free of the spectrum of the pencils $\mathfrak
A^1,\dots,\mathfrak A^N$. Assume that $T$ is sufficiently large.
Then the operator (\ref{o1}) of the problem (\ref{1}) is Fredholm.
\end{prop}

Let us introduce the space of waves $\mathcal W_\gamma (G)$.
Suppose that the assumptions of Proposition~(\ref{pF}) are
fulfilled. We extend the functions $\chi z^\pm_j$,
$j=1,\dots,M^r$, from the semicylinder $\Pi^r_+$ to the domain $G$
by zero and set
\begin{equation}\label{wv}
v^\pm_h:=\chi z^\pm_j ,\quad h=j+\sum_{p=1}^{r-1}M^p, \
j=1,\dots,M^r,\ r=1,\dots,N;
\end{equation}
 here $z^\pm_j$ are defined by (\ref{3.5}), the cut-off function
 $\chi$ is the same as in Proposition~\ref{p3.3}.
  Let $\mathcal W_\gamma (G)$ be the space spanned by functions
of the form $v^\pm_h+v$, where $v$ is a function in $\mathcal
D^\ell_{\gamma}(G)$, and $h=1,\dots,M$, $M=\sum_{r=1}^N M^r$. It
is clear that $\mathcal W_\gamma (G)\subset \mathcal
D_{-\gamma}^\ell (\Pi^r)$.  Note that the elements of $\mathcal
W_\gamma(G)$ do not necessary satisfy the homogeneous
problem~(\ref{1}).

Denote
\begin{equation}\label{q}
q(u,v)=(\mathcal Lu,v)_G+(\mathcal B u, \mathcal Q v)_{\partial
G}- (u,\mathcal L v)_G-(\mathcal Q u, \mathcal B v)_{\partial G}.
\end{equation}
The quantity  $iq ( u, u)$ represents the total energy flow
transferred by the wave $u\in\mathcal W_\gamma(G)$ through the
infinitely distant cross-sections $\Omega^1, \dots, \Omega^N$ of
the cylindrical ends of the domain $G$. It is easy to see that $iq
( u, u)=0$ for an exponentially decreasing function $u\in \mathcal
D_\varepsilon^l (G)$, $\varepsilon>0$.
\begin{lm}\label{pocv} Under the circumstances of Proposition~\ref{pF} the
waves $ v^\pm_j$, $j=1,\dots,M$, given by (\ref{wv}) satisfy the
conditions
\begin{equation}\label{ocv}
q(v^\pm_h,v^\pm_j)=\mp i \delta_{h,j},\quad q(v^\pm_h,v^\mp_j)=0,
\quad j,h=1,\dots,M.
\end{equation}
Thus $v_1^+,\dots,v_{M}^+$ are incoming waves and $v_1^-,\dots,
v_{M}^-$ are outgoing waves for the problem~(\ref{1}).
\end{lm}
\noindent{\bf Proof.} By Proposition~\ref{p3.3} the conditions
(\ref{ocz}) are valid. Due to the Green formula (\ref{GFg}) we can
replace in (\ref{ocz}) the cut-off function $\chi$ by a cut-off
function $\zeta_T\in C^\infty (\Bbb R)$, $\zeta_T(t)=1$ for $t>3T$
and $\zeta_T(t)=0$ for $t<2T$. Recall that the operator
$\{\mathcal L,\mathcal B\}$ coincides with $\{\mathfrak
L^r_T,\mathfrak B^r_T\}$ on the set
$\{(y^r,t^r)\in\overline\Pi^r:t^r>T+3\}$; see section~\ref{sGFg}.
 If  $v^\pm_j$ and $v^\pm_h$ are related to different
semicylinders $\Pi^r_+$ and $\Pi^s_+$ then those supports do not
overlap. We have
$$
q(\zeta_T v^\pm_h,\zeta_T v^\pm_j)=\mp i \delta_{h,j},\quad
q(\zeta_T v^\pm_h,\zeta_T v^\mp_j)=0, \quad j,h=1,\dots,M.
$$
Owing to the Green formula (\ref{GF}) the cut-off function
$\zeta_T$ can be omitted. \qed

\begin{thm}\label{pasv} Let $\mathcal L(x,D_x)$ and $\mathcal R(x,D_x)$
stabilize in $\Pi_+^1,\dots,\Pi_+^N$ and let the line $\Bbb R+
i\gamma$ be free of the spectrum of the pencils $\mathfrak
A^1,\dots,\mathfrak A^N$. Then for a solution $u\in \mathcal
D^\ell_{-\gamma}(G)$ to the problem~(\ref{1}) with right-hand side
$\{\mathcal F,\mathcal G\}\in \mathcal R_\gamma^\ell (G)$ the
inclusion
\begin{equation}\label{asv}
u-\sum_{j=1}^M\{a_j v_j^++b_j v_j^-\}\in\mathcal
D^\ell_{\gamma}(G)
\end{equation}
holds. Here
\begin{equation}\label{cv}
a_j=iq(u,v^+_j), \quad b_j=-iq(u,v^-_j), \quad j=1,\dots, M,
\end{equation}
 the waves $v^\pm_j$ are defined by (\ref{wv}) and
(\ref{3.5}), where $T$ is sufficiently large.
\end{thm}

\noindent{\bf Proof.} Let $\zeta_T\in C^\infty(\overline G)$,
$\zeta_T(t)=1$ for $t>3T$ and $\zeta_T(t)=0$ for $t<2T$. Denote by
$\zeta_T^r$ the cut-off function such that $\zeta_T^r$ coincides
with $\zeta_T$ inside  $\overline\Pi^r_+$ and vanishes on the
remaining part of $\overline G$. Theorem~\ref{mp} implies the
representations of the form (\ref{asz}) for the solutions
$\zeta^r_T u\in \mathcal D_{-\gamma}^l(\Pi^r)$ to the problems
(\ref{pp}) with the right-hand sides $\{\mathfrak F^r, \mathfrak
G^r\}:=\{\mathfrak L^r_T,\mathfrak B^r_T\}\zeta^r_T u$,
$r=1,\dots, N$. To prove (\ref{asv}) it remains to note that
$\zeta_T\chi=\zeta_T$ and $(1-\sum_{r=1}^N\zeta^r_T)u\in \mathcal
D^\ell_\gamma (G)$.

The equalities (\ref{cv}) directly follow from (\ref{asv}) and
Lemma~\ref{pocv}.\qed

\begin{thm}\label{taswG}
Assume that $\mathcal L$ and $\mathcal R$ stabilize in $\Pi^r_+$.
We also suppose that $\beta>\alpha$ and the lines $\Bbb R+i\alpha$
and $\Bbb R+i\beta$ contain no eigenvalues of the pencil
$\mathfrak A^r$. Let $\lambda_K,\dots,\lambda_M$ be all
eigenvalues of $\mathfrak A^r$ from the strip
$\alpha<\operatorname{Im}\lambda<\beta$ and let $\eta \{\mathfrak
F,\mathfrak G\}\in \mathcal R^l_\beta(G)$, where $\eta\in C^\infty
(\overline G)$, $\operatorname{supp} \eta\in \overline\Pi^r_+$ and
$\eta=1$ on the set $\{(y,t)\in\overline\Pi^r_+, t>3\}$. If $u$ is
a solution to the problem (\ref{1}) such that $\eta u \in \mathcal
D_\alpha^l(G)$ then inside $\Pi^r_+$ the representation
\begin{equation}\label{str}
u=\sum_{\nu=K}^M\sum_{j=1}^{J_\nu}\sum_{\sigma=0}^{\varkappa_{j\nu}-1}
c_\nu^{(\sigma,j)}(\mathfrak F,\mathfrak G)\,w_\nu^{(\sigma,j)} +v
\end{equation}
holds, where $\eta v\in \mathcal D_\beta^l(G)$ and
\begin{equation*}
c_\nu^{(\sigma,j)}(\mathfrak F,\mathfrak G) =i q(u, \eta
w_{-\nu}^{(\varkappa_{j\nu}-\sigma-1,j)}),\quad |\nu|>\nu_0,
\end{equation*}
\begin{equation*}
 c_\nu^{(\sigma,j)}(\mathfrak F,\mathfrak G)=\pm i q(u, \eta
w_{\nu}^{(\varkappa_{j\nu}-\sigma-1,j)}),\quad |\nu|\leq\nu_0.
\end{equation*}
The sign in the last formula is the same as in~(\ref{2.14}).
\end{thm}
\noindent{\bf Proof.} The assertion follows from the item (i) of
Theorem~\ref{tasw} and Proposition~\ref{bang-up}.
 \qed
\section{Corollaries of Theorems~\ref{pasv} and \ref{taswG}}

\subsection{Index properties, Scattering matrices, An existence criterion of exponentially decaying solutions}
\begin{prop}\label{cl1} Let the assumptions of Theorem~\ref{taswG} be fulfilled.
Then the indexes of operators ${\mathcal A}(\alpha)$ and
${\mathcal A}(\beta)$ are connected by the relation
$$\operatorname{Ind}{\mathcal A}(\alpha)-\operatorname{Ind}{\mathcal A}(\beta)=\varkappa,
$$
where $\varkappa$ is the total algebraic multiplicity of all
eigenvalues of the pencils ${\mathfrak A}^1,\ldots,{\mathfrak
A}^N$ in the strip $\{\lambda\in{\mathbb C}:
\alpha<\operatorname{Im}\lambda<\beta\}$.
\end{prop}
The assertion of this proposition follows from the structure
(\ref{str}) of solution to the problem (\ref{1}); see
\cite[Section 4.3]{3}. Using Proposition~\ref{cl1} and the formal
self-adjointness of $\{\mathcal L,\mathcal B\}$  one can prove the
following proposition; see \cite[Section 5.1.3]{3}.
 \begin{prop}\label{cl2}
Let the assumptions of Theorem~\ref{pasv} be fulfilled. Then
\begin{equation*}
 \operatorname{dim}\operatorname{ker}{\mathcal
A}(-\gamma)-\operatorname{dim}\operatorname{ker}{\mathcal
A}(\gamma) =\operatorname{dim}\operatorname{coker}{\mathcal
A}(\gamma)-\operatorname{dim} \operatorname{coker}{\mathcal
A}(-\gamma)=M,
\end{equation*}
where $2M$ is the total algebraic multiplicity of all eigenvalues
of the pencils ${\mathfrak A}^1,\ldots,{\mathfrak A}^N$ in the
strip $\{\lambda\in{\mathbb C}:
|\operatorname{Im}\lambda|<\gamma\}$.
\end{prop}

The next proposition is a corollary of the formulas (\ref{cv}) for
the coefficients in the structure (\ref{asv}) of solution; see
\cite[Propositions 5.3.3, 5.3.4]{3}.
\begin{prop}\label{cl5} Let the assumptions of Theorem~\ref{pasv} be fulfilled. Then there exist  bases $Z_1,\ldots,Z_M$ and
$X_1,\ldots,X_M$ in the space $\operatorname{ker} {\mathcal
A}(-\gamma)$
 modulo ${\mathcal
D}_\gamma^\ell W(G)$ such that
\begin{eqnarray}
Z_k-\biggl(v_k^++\sum_{j=1}^M {\mathfrak T}_{kj} v_j^-\biggr)\in
{\mathcal D}_\gamma^\ell W(G),\quad k=1,\ldots,M,\label{b1}\\
X_k-\biggl(v_k^-+\sum_{j=1}^M {\mathfrak S}_{kj} v_j^+\biggr)\in
{\mathcal D}_\gamma^\ell W(G),\quad k=1,\ldots,M,\label{b2}
\end{eqnarray}
where the scattering  matrices ${\mathfrak T}\equiv\|{\mathfrak
T}_{kj}\|$ and ${\mathfrak S}\equiv\|{\mathfrak S}_{kj}\|$ of
sizes $M\times M$ are unitary, i.e. ${\mathfrak T}^*={\mathfrak
T}^{-1}$ and ${\mathfrak S}^*={\mathfrak S}^{-1}$; moreover,
${\mathfrak S}={\mathfrak T}^{-1}$.
\end{prop}

Before formulating an existence criterion of exponentially
decaying solutions to the homogeneous problem~(\ref{1}) we need to
construct a special basis $\{v^\pm_j\}_{j=1}^{M'}$ modulo
$\mathcal D_\beta^l (G)$ in the space of waves $\mathcal
W_\beta(G)$, $\beta>\gamma$.
\begin{lm}\label{lm} Let $0<\gamma<\beta$, and let  $\{v^\pm_j\}_{j=1}^M$
 be a basis  in the space of
 waves $\mathcal W_{\gamma}(G)$ modulo $\mathcal D_{\gamma}^l(G)$ subjected to (\ref{ocv}).
The set $\{v^\pm_j\}_{j=1}^M$ can be supplemented to a basis
$\{v^\pm_j\}_{j=1}^{M'}$ in $\mathcal W_{\beta}(G)$ modulo
$\mathcal D_{\beta}^l(G)$ so that $v_s^++v_s^-\in {\mathcal
D}_\gamma^l W(G)$ for $s=M+1,\ldots,M'$ and the relations
{\rm(\ref{ocv})} hold for $h,j=1,\ldots,M'$.
\end{lm}

\noindent{\bf Proof.} In fact the waves $v^\pm_s$, $s=M+1,\dots,
M'$ can be constructed in the same way as the waves $u^\pm_s$ (see
e.g. \cite{11}), one has to use the functions (\ref{w}) instead of
functions (\ref{2.11}).

 For simplicity of description we suppose
that domain $G$ has only one cylindrical end $\Pi^1_+$. Assume
that the Jordan chains of the pencil $\mathfrak A^1$ are chosen
such that the conditions (\ref{2.13})--(\ref{2.15}) for the
functions (\ref{2.11}) are valid. To the every Jordan chain
$\{\varphi_\nu^{(0,j)},\dots,\varphi_\nu^{(\varkappa_{j\nu}-1,j)}\}$
there correspond the functions
$w_\nu^{(0,j)},\dots,w_\nu^{(\varkappa_{j\nu}-1,j)}$ given by
(\ref{w}). By Proposition~\ref{bang-up} the conditions
(\ref{w.1})--(\ref{w.3}) are valid. With every
 eigenvalue $\lambda_\nu$ of $\mathfrak A^1$ such that
$\gamma<\operatorname{Im}\lambda_\nu<\beta$ we associate the
functions
\begin{equation}\label{002}
w_{\nu,\pm}^{(\sigma,j)}=2^{-1/2}\chi (w_\nu^{(\sigma,j)}\mp
w_{-\nu}^{(\varkappa_{j\nu}-\sigma-1,j)}),\quad j=1,\dots, J_\nu,
\tau=0,1,\dots,\varkappa_{j\nu}-1,
\end{equation}
where $J_\nu=\dim \ker \mathfrak A^1(\lambda_\nu)$, $\chi$ is the
same as in (\ref{wv}). Then owing to (\ref{w.1}) and (\ref{w.3})
we have
$p^1_T(w_{\nu,\pm}^{(\sigma,j)},w_{\mu,\pm}^{(\tau,p)})=\mp
i\delta_{\nu,\mu}\delta_{\sigma,\tau}\delta_{j,p}$; the wave
$w_{\nu,+}^{(\sigma,j)}$ is incoming and the wave
$w_{\nu,-}^{(\sigma,j)}$ is outgoing.
 Due to
the linear independence of the functions $w_\nu^{(\sigma,j)}$ and
Lemma~\ref{pwz}, the elements of the basis $\{v^\pm_j\}_{j=1}^M$
can be expressed in terms of functions $\chi w_\nu^{(\sigma,j)}$
corresponding to the eigenvalues of $\mathfrak A^1$ in the strip
$\{\lambda\in\Bbb C: |\operatorname{Im}\lambda|<\gamma\}$.
Together with (\ref{w.3}) this implies
$p^1_T(w_{\nu,\pm}^{(\sigma,j)}, v^\pm_s)=0$ for $s=1,\dots,M$. It
remains to note that
$w_{\nu,+}^{(\sigma,j)}+w_{\nu,-}^{(\sigma,j)}=2\chi
w_\nu^{(\sigma,j)}\in \mathcal D_\gamma^l(G)$. As $v^\pm_j$,
$j=M+1,\dots,M'$, we can take the waves
$w_{\nu,\pm}^{(\sigma,j)}$. \qed

\begin{prop}\label{crit}
Let $0<\gamma<\beta$, and let the lines ${\mathbb R}+i\gamma$ and
${\mathbb R}+i\beta$ be free of the spectrum of the pencils
${\mathfrak A}^1,\ldots,{\mathfrak A}^N$. Denote by $\mathfrak
S=\mathfrak S(\beta)$ the scattering matrix corresponding to the
basis $\{v_j^\pm\}_{j=1}^{M'}$ from Lemma~\ref{lm}. Then
$$
\operatorname{dim} \operatorname{ker} {\mathcal
A}(\gamma)-\operatorname{dim}\operatorname{ ker}{\mathcal
A}(\beta)=\operatorname{dim} \operatorname{ker}({\mathfrak
S}^{2,2}-I),
$$
where ${\mathfrak
 S}^{2,2}$ is  $(M'-M)\times (M'-M)$-block of the
$M'\times M'$-matrix $\mathfrak S=\|{
S}^{k,\ell}(\gamma')\|_{k,\ell=1,2}$.
\end{prop}
The proof is similar to the proof of Theorem~3.3 from~\cite{11}.

\subsection{Problem with radiation conditions}
As before we suppose that $\mathcal L(x,D_x)$ and $\mathcal
R(x,D_x)$ are  stabilizing in $\Pi_+^1,\dots,\Pi_+^N$ and  the
line $\Bbb R+ i\gamma$ is free of the spectrum of the pencils
$\mathfrak A^1,\dots,\mathfrak A^N$. Denote by $\mathcal
W_{out}(G)$ the linear span of the outgoing waves
$v^-_1,\dots,v^-_M$ and consider the restriction $A$ of $\mathcal
A(-\gamma)$ to the space $\mathfrak D_{out}(G)=\mathcal
W_{out}(G)[\dot{+}]\mathcal D_\gamma^l(G)$, where by $[\dot{+}]$
we denote the orthogonal with respect to the form (\ref{q}) direct
sum. The mapping $A:\mathfrak D_{out}(G)\to\mathcal R_\gamma^l(G)$
is continuous.

\begin{prop}\label{cl3}
Let $z_1, \dots, z_d$ be a basis of $\ker \mathcal A(\gamma)$, and
let $\{f,g\}\in\mathcal R_\gamma^l(G)$, $(f,z_j)_G+(g,\mathcal Q
z_j)_{\partial G}=0$, $j=1,\dots,d$.

\noindent{\rm (i)}~There exists a unique up to an arbitrary
element of $\operatorname{ker}{\mathcal A}(\gamma)$ solution
$u\in\mathfrak D_{out}(G)$ to the problem (\ref{1}).

\noindent{\rm (ii)}~The  inclusion
\begin{equation*}
v\equiv u-b_1v^-_1-b_2v^-_2-\ldots-b_M v^-_M\in{\mathcal
D}_\gamma^\ell W(G)
\end{equation*}
holds with the coefficients
$$
b_j=-i(f,X_j)_G-i(g,{\mathcal Q} X_j)_{\partial G}, \quad
j=1,\dots, M,
$$
where $X_1,\dots,X_M$ are elements of $\ker\mathcal A(-\gamma)$
subjected to (\ref{b2}).

\noindent{\rm (iii)}~The solution $u$ satisfies the inequality
\begin{equation}\label{e1}
\begin{split}
 \|v;{\mathcal D}_\gamma^\ell W(G)\|&+|b_1|+|b_2|+\ldots+|b_M|
\\&\leqslant C(\|\{f,g\};{\mathcal R}^\ell_\gamma W(G)\|+\|e_\gamma
v;L_2(G)\|).
\end{split}
\end{equation}

{\rm 4.}~The solution $u$ subjected to the additional conditions
$(u,z_j)_G=0$, $j=1,\ldots,d$, is unique and satisfies the
estimate {\rm(\ref{e1})} with the right-hand side replaced by
$\|\{f,g\}; {\mathcal R}^\ell_\gamma W(G)\|$.
\end{prop}
This proposition justifies the statement of the problem (\ref{1})
with intrinsic radiation conditions {\rm(}only outgoing
{\rm``}waves{\rm''} occur in asymptotic formulas for
solutions{\rm)}. Up to obvious changes the proof repeats the proof
of Theorem 5.3.5 from \cite{3}.  The next two propositions
describe the statement of the problem with other radiation
conditions. For the proofs we refer to \cite[Theorems 5.5.5,
5.5.6]{3}.
\begin{prop}\label{cl4}
Let $\eta_1,\ldots,\eta_{d}$ be a basis of $\operatorname{ker}
{\mathcal A}(\gamma)$, and let the right-hand side $\{f,g\}\in
{\mathcal R}_\gamma^\ell W(G)$ satisfy the orthogonality
conditions $(f,\eta_j)_G+(g, {\mathcal Q} \eta_j)_{\partial G}=0$,
$j=1,\ldots,d$. We assume that for the space $\operatorname{ker}
{\mathcal A}(-\gamma)$ one can choose a basis $V_1,\ldots,V_M$
modulo ${\mathcal D}_\gamma^l W(G)$ that compatible with the basis
$u_1,\ldots,u_{2M}$ for the quotient space ${\mathcal
W}_\gamma(G)/{\mathcal D}_\gamma^\ell W(G)$ in the following
sense{\rm:}
\begin{equation}
q(u_j,V_k)=-i\delta_{kj}, \quad k,j=1,\ldots,M. \label{ccc}
\end{equation}
Then the following assertions hold.

{\rm 1.}~There exists a unique up to an arbitrary element of
$\operatorname{ker}{\mathcal A}(\gamma)$ solution $u\in {\mathfrak
h[\dot +}]{\mathcal D}_\gamma^\ell W(G)$ to the problem {\rm
(\ref{1})}, where $ {\mathfrak h}$ is the linear span of the
functions $u_1, \ldots,u_M$ .

{\rm 2.}~The following inclusion holds{\rm:}
\begin{equation*}
v\equiv u-b_1u_1-b_2u_2-\ldots-b_Mu_M\in{\mathcal D}_\gamma^\ell
W(G),
\end{equation*}
where $b_j=i(f,V_j)_G+i(g,{\mathcal Q} V_j)_{\partial G}$,
$j=1,\ldots, M$.

{\rm 3.}~The solution $u$ satisfies the inequality~(\ref{e1}).

{\rm 4.} The solution $u$ subjectd to the additional conditions
$(u,\eta_j)_G=0$, $j=1,\ldots,d$, is unique and satisfies the
estimate {\rm(\ref{e1})} with the right-hand side replaced by
$\|\{f,g\}; {\mathcal R}^\ell_\gamma W(G)\|$.
\end{prop}
By Proposition \ref{cl4}, to enumerate all possible radiation
conditions is the same that to enumerate all the bases
$u_1,\ldots,u_{2M}$ for the quotient space ${\mathcal
W}_\gamma(G)/ {\mathcal D}_\gamma^\ell W(G)$ and bases
$V_1,\ldots,V_M$ modulo ${\mathcal D}_\gamma^\ell W(G)$ for the
subspace $\operatorname{ker}{\mathcal A}_*(-\gamma)$ compatible in
the sense of (\ref{ccc}).

\begin{prop} \label{cl6}
Let ${\mathcal W}_\gamma(G)$ be the space of waves and let the
waves $v^\pm_j$, $j=1,\dots,M$, form a basis of the quotient space
${\mathcal W}_\gamma(G)/{\mathcal D}_\gamma^\ell W(G)$ subjected
to (\ref{ocv}). Denote by $\{X_1,\ldots,X_M\}$  a set of solutions
to the  homogeneous problem $\{{\mathcal L},{\mathcal B}\}u=0$
satisfying the inclusions {\rm(\ref{b2})}. Then the following
assertions hold.

{\rm 1.}~If $R$ is arbitrary and $S$ is an invertible operator in
${\mathbb C}^M$, then
\begin{equation}
\begin{aligned}
{}&V_k=\sum_{m=1}^M \overline{(S^{-1})_{mk}} X_m,\\
&u_j=\sum_{m=1}^M \biggl(S_{jm}u_m^-+\sum_{p=1}^M
R_{jp}\biggl\{u_p^++\sum_{i=1}^M{\mathfrak S}^*_{pi}u_i^-
\biggr\}\biggr),
\end{aligned}
\label{++++}
\end{equation}
where $j,k=1,\ldots,M$, satisfy the condition {\rm(\ref{ccc})}.

{\rm 2.}~If a basis $V_1,\ldots,V_M$ modulo ${\mathcal
D}_\gamma^\ell W(G)$ for the space $\operatorname{ker}{\mathcal
A}(-\gamma)$ and a basis $u_1,\ldots,u_{2M}$ for the quotient
space ${\mathcal W}_\gamma(G)/{\mathcal D}_\gamma^\ell W(G)$
satisfy {\rm(\ref{ccc})}, then there exist operators $R$ and $S$
such that the equalities {\rm(\ref{++++})} hold.
\end{prop}
\subsection{The extensions of the symmetric operator}\label{sss}
The schemes for the proofs of propositions listed in this section
can be found in \cite[Section 5.5]{3}, the changes in the proofs
consist in usage of Theorem~\ref{pasv} instead of asymptotic
representations.

Here we assume that the elliptic system $\{\mathcal L,\mathcal
B\}$ is homogeneous. In other words,
$\tau_1=\tau_2=\ldots=\tau_k\equiv \tau$ and $ {\mathcal
D}_\gamma^\ell W(G):=\prod_{i=1}^k W_\gamma^{2\tau}(G)$. With the
problem ({2.1}) we associate an operator ${\mathcal M}$ with the
domain
$$
{\mathcal D}({\mathcal M})=\{u\in{\mathcal D}_\gamma^\ell W(G):
{\mathcal B}(x,D_x)u(x)=0, x\in\partial G\}
$$
that acts in the Hilbert space
$$
L_2(G;e_{-\gamma})\equiv \prod_{i=1}^k W_{-\gamma}^0 (G)
$$
by the formula
\begin{equation*}
({\mathcal M} u)(x)=e_\gamma(x)^2{\mathcal L}(x,D_x)u(x).
\end{equation*}
We denote by $(\cdot,\cdot)_{-\gamma}$ the inner product
\begin{equation*}
(u,v)_{-\gamma}=\int\limits_G e_{-\gamma}(x)^2
u(x)\overline{v(x)}\, dx
\end{equation*}
in the space $L_2(G;e_{-\gamma})$.

\begin{prop}
Suppose that $\mathcal L(x,D_x)$ and $\mathcal R(x,D_x)$ are
stabilizing and the line ${\mathbb R}+i\gamma$ is free of the
spectrum of the pencils $ {\mathfrak A}^1,\ldots,{\mathfrak A}^N$.
Then the operator ${\mathcal M}$ is closed and symmetric. For the
$ \operatorname{coker} {{\mathcal M}}$ one can choose a basis
$X_1,\ldots,X_M$ modulo ${\mathcal D}_\gamma^l W(G)$ such that the
inclusions {\rm(\ref{b2})} hold.
\end{prop}

\begin{prop}

Let $\mathcal L(x,D_x)$ and $\mathcal R(x,D_x)$ stabilize and let
the line ${\mathbb R}+i\gamma$ be free of the spectrum of the
pencils $ {\mathfrak A}^1,\ldots,{\mathfrak A}^N$. The operator
${\mathcal M}^*$, adjoint to ${\mathcal M}$ in the space
$L_2(G;e_{- \gamma})$, is defined on the set
\begin{equation*}
{\mathcal D}({\mathcal M}^*)=\{v\in{\mathcal W}_\gamma(G):
{\mathcal B}(x,D_x)v(x)=0,x\in\partial G\} \label{5.5}
\end{equation*}
and acts by the formula $({\mathcal M}^*
u)(x)=e_\gamma(x)^2{\mathcal L}(x,D_x)u(x)$.
\end{prop}

 We extend the operator ${\mathcal M}$  by adjoining representatives of waves to the domain.
Let $u\in\mathcal D({\mathcal M}^*)$. Then $u$ is a solution of
the problem~(\ref{1}) with the right-hand side $\{f,g\}=\{\mathcal
L u, 0\}\in L_2(G;e_\gamma)$. In accordance with
propositions~\ref{cl4} and~\ref{cl5} the function $u$ has the form
\begin{equation*}
u=u^0+\sum_{k=1}^d c_k \eta_k+\sum_{j=1}^M d_j X_j,
\end{equation*}
where $u^0$ is such that $u^0-b_1 v^-_1-\cdots-b_M v^-_M\in
\mathcal D^l_\gamma (G)$.  With every function $u\in\mathcal
D({\mathcal M}^*)$ we associate vectors $d=(d_1,\dots,d_M)$ and
$b=(b_1,\dots,b_M)$ of the coefficients.
\begin{prop}
Suppose that $\mathcal L(x,D_x)$ and $\mathcal R(x,D_x)$ are
stabilizing and the line ${\mathbb R}+i\gamma$ is free of the
spectrum of the pencils $ {\mathfrak A}^1,\ldots,{\mathfrak A}^N$.
Let $\Bbb C^M=\Bbb N_+(+)\Bbb N_0(+)\Bbb N_-$ be an orthogonal sum
of subspaces and let $K$ be a self-adjoint operator in $\Bbb N_0$.
The operator $\Bbb M$ is a self-adjoint extension of $\mathcal M$
if and only if $\Bbb M$ is defined on the set
\begin{equation*}
\begin{split}
\mathcal D(\Bbb M)=\{&u\in\mathcal D (\mathcal M^*):
c=\zeta_++\zeta_0,\\&
a=\zeta_--(iK+I)\zeta_0/2-\zeta_+/2,\zeta_\pm\in\Bbb N_\pm,
\zeta_0\in\Bbb N_0\}
\end{split}
\end{equation*}
 and acts by the formula $({\Bbb M} u)(x)=e_\gamma(x)^2{\mathcal L}(x,D_x)u(x)$.
\end{prop}

\end{document}